\documentclass[traditabstract]{aa} 
\usepackage{graphicx}
\usepackage{txfonts}
\usepackage{natbib}
\newcommand{\kms}{{\hbox {km\thinspace s$^{-1}$}}}
\newcommand{\Lsun}{{\hbox {L$_\odot$}}}
\newcommand{\Msun}{{\hbox {M$_\odot$}}}
\newcommand{\Mdot}{{\hbox {$\dot{M}$}}}

\newcommand{\cmt}{{\hbox {cm$^{-3}$}}}
\newcommand{\cmd}{{\hbox {cm$^{-2}$}}}
\newcommand{\hdo}{{\hbox {H$_{2}$O}}}

\def\t#1#2#3#4#5#6{{\hbox {$#1_{#2#3}-#4_{#5#6}$}}}

\def\13co{$^{13}$CO}
\def\c18o{C$^{18}$O}

\newcommand{\tdust}{{\hbox {$T_{\mathrm{dust}}$}}}

\newcommand{\cext}{{\hbox {$C_{\mathrm{extended}}$}}}
\newcommand{\chalo}{{\hbox {$C_{\mathrm{halo}}$}}}

\begin{document}
   \title{The Mrk 231 molecular outflow as seen in OH\thanks{Herschel is an
       ESA space observatory with science 
       instruments provided by European-led Principal Investigator consortia
       and with important participation from NASA.} }

   \author{E. Gonz\'alez-Alfonso \inst{1}, J. Fischer \inst{2},
     J. Graci\'a-Carpio \inst{3}, N. Falstad \inst{4}, E. Sturm \inst{3},
     M. Mel\'endez \inst{5}, H. W. W. Spoon \inst{6}, A. Verma \inst{7},
     R. I. Davies \inst{3}, D. Lutz \inst{3}, S. Aalto \inst{4}, E. Polisensky
     \inst{2}, A. Poglitsch \inst{3}, S. Veilleux \inst{5}, 
     A. Contursi \inst{3} 
}

   \institute{Universidad de Alcal\'a, Departamento de F\'{\i}sica
     y Matem\'aticas, Campus Universitario, E-28871 Alcal\'a de Henares,
     Madrid, Spain  
         \and
Naval Research Laboratory, Remote Sensing Division, 4555
     Overlook Ave SW, Washington, DC 20375, USA
         \and
Max-Planck-Institute for Extraterrestrial Physics (MPE), Giessenbachstra{\ss}e
1, 85748 Garching, Germany 
         \and
Department of Earth and Space Sciences, Chalmers University of Technology,
Onsala Space Observatory, Onsala, Sweden 
         \and
Department of Astronomy, University of Maryland, College Park, MD 20742, USA
         \and
Cornell University, Astronomy Department, Ithaca, NY 14853, USA
         \and
University of Oxford, Oxford Astrophysics, Denys Wilkinson Building, Keble
Road, Oxford, OX1 3RH, UK 
}


 
   \authorrunning{Gonz\'alez-Alfonso et al.}
   \titlerunning{The Mrk 231 molecular outflow as seen in OH}

  \abstract
   { We report on the Herschel/PACS observations of OH in Mrk~231, with
     detections in nine doublets observed within the PACS range, and
     present radiative-transfer models for the outflowing OH. Clear
     signatures 
     of outflowing gas are found in up to six OH doublets with different
     excitation requirements. At least two outflowing components are
     identified, one with OH radiatively excited, and the other with low
     excitation, presumably spatially extended and roughly spherical.
     Particularly prominent, the blue wing of the absorption detected in the
     in-ladder $^{2}\Pi_{3/2}\,J=9/2-7/2$ OH doublet at 65 $\mu$m, with
     $E_{\mathrm{lower}}=290$ K, indicates that the excited outflowing gas is
     generated in a compact and warm (circum)nuclear region. Because the
     excited, outflowing OH gas in Mrk~231 is associated with the warm,
     far-infrared continuum source, it is most likely more compact
     (diameter of $\sim200-300$ pc) than that probed by CO and
     HCN. Nevertheless, its mass-outflow rate per unit of solid angle as 
     inferred from OH is similar to that previously derived from CO,
     $\gtrsim70\times(2.5\times10^{-6}/X_{\mathrm{OH}})$ \Msun\ yr$^{-1}$
     sr$^{-1}$, where $X_{\mathrm{OH}}$ is the OH abundance relative to H
       nuclei. In spherical symmetry, this would correspond to
     $\gtrsim850\times(2.5\times10^{-6}/X_{\mathrm{OH}})$ 
     \Msun\ $\mathrm{yr^{-1}}$, though significant collimation is inferred
     from the line profiles. The momentum flux of the excited
       component attains $\sim15\,L_{\mathrm{AGN}}/c$, 
     with an OH column density of $(1.5-3)\times10^{17}$
     cm$^{-2}$ and a mechanical luminosity of $\sim10^{11}$ \Lsun.  
     In addition, the  detection of very excited, radiatively pumped OH
     peaking at central velocities indicates the presence of a nuclear
     reservoir of gas rich in OH, plausibly the 130-pc scale circumnuclear
     torus previously detected in OH megamaser emission, that may be
     feeding the outflow. An exceptional $^{18}$OH enhancement, with
     $\mathrm{OH/^{18}OH}\lesssim30$ at both central and blueshifted
     velocities, is most likely the result of interstellar-medium
     processing by recent starburst and supernova activity within the
     circumnuclear torus or thick disk.}


   \maketitle
%

\section{Introduction}

Current models of galaxy evolution involve galactic-scale outflows driven
by starbursts and active galactic nuclei (AGN) as key ingredients. The
outflows trace 
the negative feedback from AGN and/or star formation on the molecular gas,
eventually shutting off the feeding process and quenching the
growth of the stellar population and/or of the supermassive black hole
\citep[e.g.][]{mat05}. AGN feedback could be responsible for the 
observed black hole mass-velocity dispersion relationship \citep{mur05}
and create a population of red gas-poor ellipticals.
In the past, outflows have been observed in many
starbursts and quasi-stellar objects (QSOs), mostly in the ionized and
neutral atomic gas component \citep[e.g.][for a review]{vei05}. 

Molecular gas may dominate the mass-outflow rate of outflows, 
providing important constraints on the timescale for dispersing
the (circum)nuclear gas in the host galaxy. 
Molecular outflows have been reported in several galaxies
at millimeter wavelengths \citep[e.g.][]{baa89,wal02,sak09}.
The discovery of a massive molecular outflow in Mrk~231, an
  ultraluminous infrared galaxy (ULIRG) harboring
the closest quasar known, in \emph{Herschel}/PACS \citep{pil10,pog10}
spectroscopy is a key finding. The outflows are traced by P-Cygni OH line
profiles in the 79 and 119 $\mu$m doublets, and in the high-lying 65 $\mu$m
doublet ($E_{\mathrm{lower}}\approx300$ K), with high-velocity shifts of
$>1000$ km/s \citep[][hereafter F10 and S11, respectively]{fis10,stu11}.   
Analysis and model fits of these lines yielded a preliminary 
  mass-outflow rate of $\dot{M}\sim10^3$ \Msun\ y$^{-1}$. The extreme 
outflow was also detected at millimeter wavelengths in CO, giving a
similar \Mdot\ \citep{fer10}, and in HCN, HCO$^+$, and HNC
\citep{aal12}. Recently, \cite{cic12} 
have found that the outflowing CO is not highly excited
relative to the quiescent gas, and that the outflow size decreases with
increasing critical density of the transition. From neutral Na I D absorption,
\cite{rup11} estimated a similar $\dot{M}\sim400$ \Msun\ y$^{-1}$ on
spatial scales of $\sim3$ kpc. 

Relative to CO, the specific characteristic of OH in galaxies is that the
high-lying lines are radiatively (instead of collisionally) excited, and thus
selectively trace an outflow region close to the circumnuclear source of
strong far-IR radiation density \citep{gon08}. In addition to tracing the
  gas, the lines also probe the coexisting warm dust responsible for the
  observed excitation. Although the
outflow is not spatially resolved, the observed excitation conditions 
provide information about the spatial extent of the outflow, which
enables the estimation of the outflow physical parameters 
(mass-outflow rate, mechanical power and energy). The high-velocity
molecular outflows were found to be common in local ULIRGs, and preliminary
evidence suggested that higher AGN luminosities (and higher AGN contributions
to $L_{\mathrm{IR}}$) correlate with higher terminal velocities and shorter
gas depletion timescales (S11).

In this work, we present the velocity profiles and fluxes of all of the
  OH and $^{18}$OH doublets seen in the \emph{Herschel}/PACS spectroscopic
  observations of Mrk 231, and an analysis of these profiles and fluxes based
  on radiative-transfer modeling. In Sect.~\ref{sec:obser} we discuss the
  details of the observations and give an overview of the general
  characteristics 
  of the profiles, together with qualitative assessments on the excitation
  conditions, optical depths, far-IR extinction, geometry, and
  $^{16}$O/$^{18}$O abundance ratio in the circumnuclear region of Mrk 231. In
  Sect.~\ref{sec:analysis}, we discuss the radiative-transfer models
  that are 
  used to quantitatively analyze the observations, and the motivation for,
  properties of, and derived parameters of the several components that
  we use to characterize the gas seen in OH. In Sect.~\ref{sec:discussion},
  we summarize the picture that emerges from the observations and the modeled
  components, their relationship to structures seen in other diagnostics, and
  the implications for the role of the AGN and circumnuclear starburst. We
  adopt a distance to Mrk 231 of 192 Mpc 
($H_0=70$ km s$^{-1}$ Mpc$^{-1}$, $\Omega_{\Lambda}=0.73$, and $z=0.04218$).


\section{Observations and description of the spectra}
\label{sec:obser}

Following the detection of outflows traced by the ground-state OH
doublets at 119 and 79 $\mu$m and of the excited OH 65 $\mu$m transition, 
based on the guaranteed-time key program SHINING observations
(hereafter GT observations, PI: E.~Sturm; F10, S11), completion of the 
full ($52.3-98$, $104.6-196$ $\mu$m), high-resolution PACS spectrum of
Mrk~231 was carried out on October 16 (2012) as part of the Open Time-2
Herschel phase (hereafter OT2 observations; PI: J.~Fischer). The spectrum of
the OH $\Pi_{1/2} \,\, J=3/2-1/2$ doublet at $163$ $\mu$m was taken from an 
OT1 program (PI: R.~Meijerink). The GT observations and reduction process
were described in F10 and S11. For the OT2 and OT1 observations, the spectra 
were also taken in high spectral sampling density mode using first and second
orders of the grating. The velocity resolution of PACS in first order ranges
from $\approx320$ \kms\ at 105 $\mu$m to $\approx180$ \kms\ at 190 $\mu$m, and
in second order from $\approx210$ \kms\ at 52 $\mu$m to $\approx110$ \kms\ at
98 microns. The data reduction was carried out mostly using the standard
PACS reduction and calibration pipeline (ipipe) included in HIPE 6.0 and
HIPE 10.0. The two HIPE versions yielded essentially identical
continuum-normalized spectra, with the continuum level from v10 stronger than
that from v6 by up to $\sim10$\%. 

   \begin{table*}
      \caption[]{Herschel/PACS observations of OH in Mrk~231.}
         \label{tab:obs}
          \begin{tabular}{lccccc}   
            \hline
            \noalign{\smallskip}
Transition  & Rest wavelengths$^{\mathrm{a}}$ & Program$^{\mathrm{b}}$ &
Obs. ID$^{\mathrm{c}}$ &  Number of  & $\sigma$ \\  
&  ($\mu$m)  &  &  & spatial pixels$^{\mathrm{d}}$ & (Jy) \\
            \noalign{\smallskip}
            \hline
            \noalign{\smallskip}
OH $\Pi_{3/2}-\Pi_{3/2}\, \frac{5}{2}-\frac{3}{2}$ & $119.233-119.441$ & GT & 1342186811 & 1 & $0.29$ \\
OH $\Pi_{1/2}-\Pi_{3/2}\, \frac{1}{2}-\frac{3}{2}$ & $79.118-79.181$ & GT  & 1342186811 & 1 & $0.26$ \\
 &  & GT   & 1342186811 & 1 & $0.26$ \\
 &  & OT2  & 1342253536 & 1 & $0.23$ \\
OH $\Pi_{1/2}-\Pi_{3/2}\, \frac{3}{2}-\frac{3}{2}$ & $53.261-53.351$ & OT2 & 1342253530 & 1 & $0.24$ \\
OH $\Pi_{3/2}-\Pi_{3/2}\, \frac{7}{2}-\frac{5}{2}$ & $84.420-84.597$ & OT2 & 1342253537 & 3 & $0.50$ \\
OH $\Pi_{3/2}-\Pi_{3/2}\, \frac{9}{2}-\frac{7}{2}$ & $65.132-65.279$ & GT &  1342207782 & 1 & $0.19$  \\
 &  & GT   & 1342207782 & 3 & $0.24$ \\
 &  & OT2  & 1342253532 & 3 & $0.25$ \\
OH $\Pi_{1/2}-\Pi_{1/2}\, \frac{7}{2}-\frac{5}{2}$ & $71.171-71.216$ & OT2 & 1342253534 & 3 & $0.36$ \\
OH $\Pi_{3/2}-\Pi_{3/2}\, \frac{11}{2}-\frac{9}{2}$& $52.934-53.057$ & OT2 & 1342253530 & 1 & $0.24$ \\
OH $\Pi_{1/2}-\Pi_{1/2}\, \frac{9}{2}-\frac{7}{2}$ & $55.891-55.950$ & OT2 & 1342253531 & 3 & $0.19$ \\
OH $\Pi_{1/2}-\Pi_{1/2}\, \frac{3}{2}-\frac{1}{2}$ & $163.124-163.397$ & OT1 & 1342223369 & 1 & $0.25$ \\
            \noalign{\smallskip}
            \hline
         \end{tabular} 
\begin{list}{}{}
\item[$^{\mathrm{a}}$] The two values correspond to the two
  $\Lambda-$components of the doublets; each one is the average of the two
  hyperfine transitions that make up a component. 
\item[$^{\mathrm{b}}$] GT and OT indicate ``guaranteed time'' and ``open
    time''.  
\item[$^{\mathrm{c}}$] Identification number. 
\item[$^{\mathrm{d}}$] Number of spatial pixels used to generate the
  spectrum. In case of 1, the central pixel was used; in case of 3, the
  spectra from the three brightest spatial pixels were coadded. In all cases,
  the absolute fluxes were scaled to match the total flux as obtained by 
  coadding the fluxes from the 25 pixels.
\end{list}
   \end{table*}

\begin{figure}
   \centering
   \includegraphics[width=7.0cm]{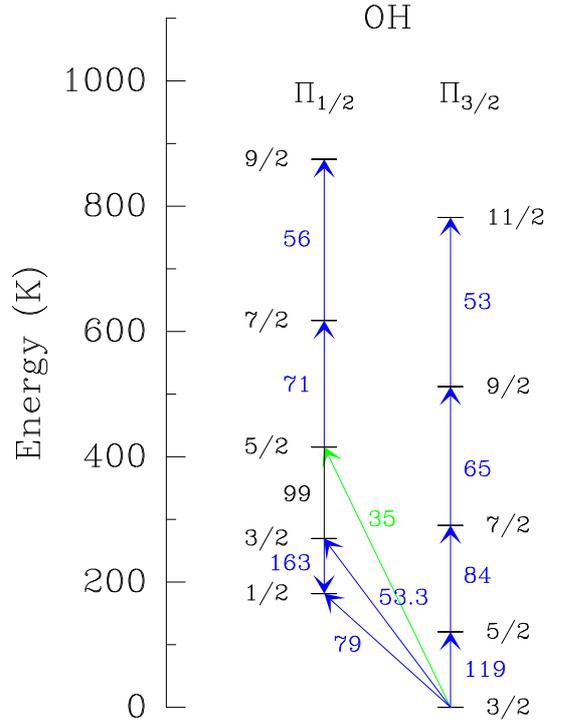}
   \caption{Energy level diagram of OH showing the transitions detected
       in Mrk~231 with Herschel/PACS and Spizer/IRS with blue and green
       arrows, respectively. The $l-$doubling splitting of the levels is 
     not indicated. Colored numbers indicate rounded
     wavelengths in $\mu$m. The $\Pi_{1/2} \,\, J=5/2-3/2$ (black line) and
     the $\Pi_{1/2}-\Pi_{3/2} \,\, J=3/2-5/2$ doublets at
     $\lambda_{\mathrm{rest}}\approx99$ and $96$ $\mu$m (the latter
       detected in NGC~4418, G-A12) were not observed, as
     they are redshifted into the PACS gap at $100$ $\mu$m. In the text, we
     denote a doublet by giving its wavelength (e.g. OH119). }     
    \label{diag}
    \end{figure}

\begin{figure*}
   \centering
   \includegraphics[width=15.0cm]{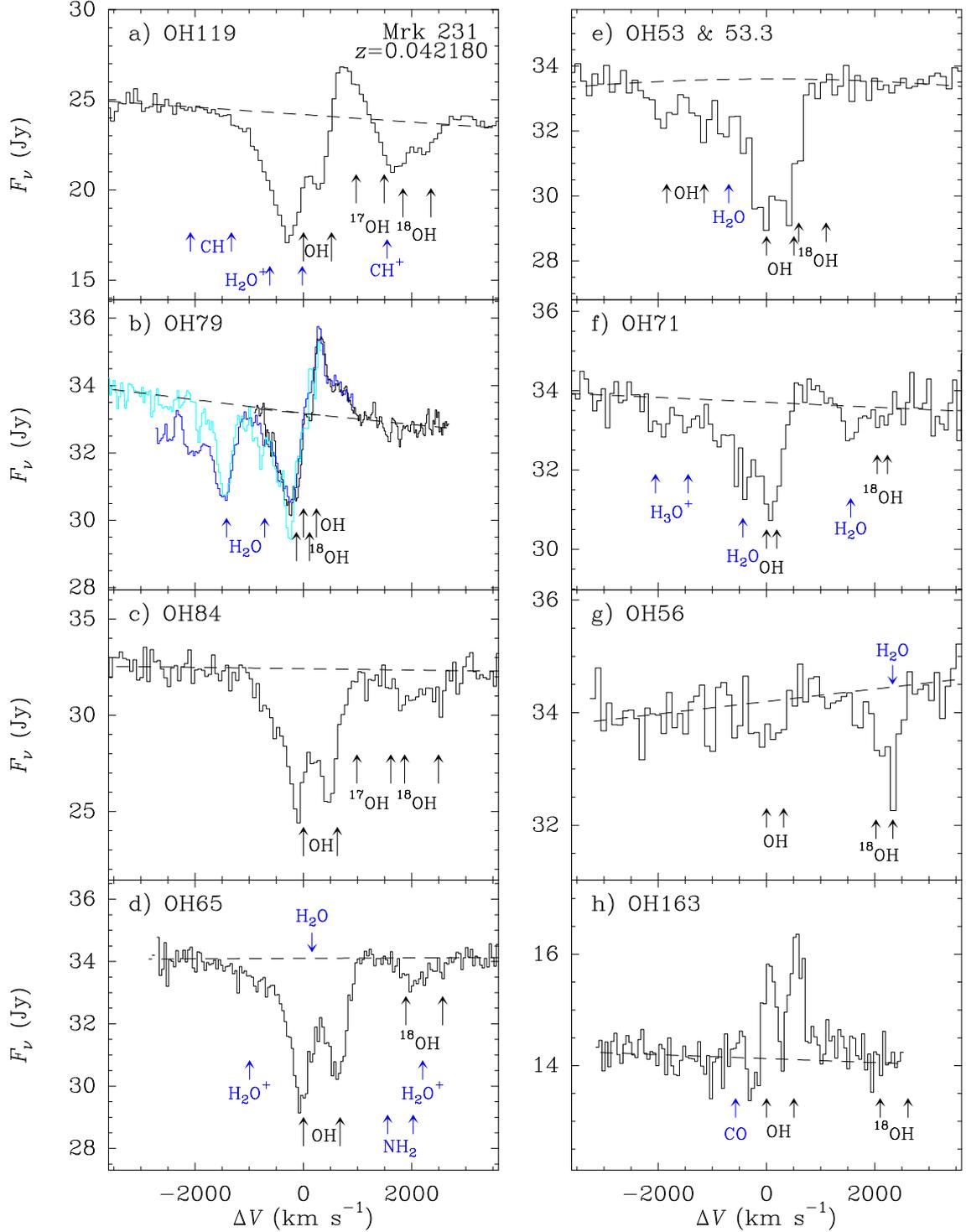}
   \caption{Herschel/PACS spectrum of Mrk~231 around the 9 OH
doublets, with the adopted baselines (dashed lines). The 
velocity scale is relative to the blue $\Lambda-$component of each
doublet and has been calculated with respect to the systemic redshift 
of $z=0.04218$. Potential contaminations due to other species are indicated
in blue (and are discussed in the text). In panel b, the black and
dark-blue OH79 spectra correspond to GT observations, while the
       light-blue spectrum corresponds to OT2 observations (see
       Table~\ref{tab:obs}). In 
       panel d, the OH65 spectrum corresponds to the average of the three
       brightest spaxels of the GT observations (GT-avg in Fig.~\ref{oh65}).
   }   
    \label{mrk231}
    \end{figure*}

We focus in this work on the observations of the OH doublets, which are
summarized in Table~\ref{tab:obs} and Fig.~\ref{diag} and are displayed
in Fig.~\ref{mrk231}. Spectroscopic parameters used for line
identification and radiative-transfer modeling were taken from the
spectral-line catalogs of the CDMS \citep{mul01,mul05} and JPL
\citep{pic98}. All nine OH doublets within the PACS wavelength range,
having lower-level energies up to $E_{\mathrm{lower}}=620$ K, were
detected and are indicated in the energy-level diagram of
Fig.~\ref{diag}. These are the same 
transitions as were detected in NGC~4418 and Arp~220 \citep[][hereafter
G-A12]{gon12}, except for the cross-ladder $\Pi_{1/2}-\Pi_{3/2} \,\,
J=3/2-5/2$ line at $\lambda_{\mathrm{rest}}=96$ $\mu$m (detected in NGC~4418)
that is redshifted in Mrk~231 into the gap at $\approx100$ $\mu$m between the
green and red bands.  
For simplicity, we denote a given doublet by using its rounded
wavelength as indicated in Fig.~\ref{diag} (e.g. OH119). The $^{18}$OH
doublets, lying close in wavelength to the OH transitions, were also observed
and unambiguously detected at 120, 85, and 65 $\mu$m. Spectroscopic
  parameters for the $^{17}$OH doublets were taken from \cite{pol03}; the
  positions of the $^{17}$OH119 and $^{17}$OH84 doublets, expected to have the
  strongest signatures, are indicated in Figs.~\ref{mrk231}a and
    c. There is no evidence in the spectra for either absorption or emission
  attributable to $^{17}$OH, whose transitions fall between those of the two
  more abundant isotopologs.

The abscissas in Fig.~\ref{mrk231} indicate the velocity relative to the
shorter-wavelength component (hereafter, blue component) of each
doublet, and are 
calculated for a redshift of $z=0.04218$. The positions of the OH and
$^{18}$OH doublets are indicated with black arrows, while those of potentially
contaminating lines of other species (discussed in Sect.~\ref{sec:cont}) are
indicated with blue arrows. In panel b, three independent spectra of the OH79
doublet are shown for comparison; they are listed in
Table~\ref{tab:obs}.   

To characterize the molecular outflow traced by the OH lines, it is
important to have a flat baseline that minimizes the uncertainties in the
continuum level, and thus in the velocity extent of the line wings. 
The central spatial pixel (spaxel) of the $5\times5$ spaxels of PACS gives the
highest signal-to-noise ratio (S/N) spectrum, and was adopted
whenever the continuum level was flat and the baseline was well
characterized. However, the continuum level from the central spaxel shows 
  low-level fluctuations at some wavelengths that probably result from
small pointing drift motions. In these 
cases, we coadded the flux densities resulting from the three brightest
spaxels, which resulted in flat baselines at the expense of a lower
S/N. Regardless of the number of spaxels used, the resulting spectra were
  re-scaled to the continuum level of all 25 spaxels combined
  (to account for point spread function losses and pointing
  uncertainties, S11). 
Table~\ref{tab:obs} lists the number of spaxels used to generate the
spectrum of each OH doublet, and the baselines are indicated in
Fig.~\ref{mrk231} with dashed lines.

 We also analyzed the OH35 ground-state doublet observed in the
  Spitzer IRS long-high spectrum of Mrk~231. The spectrum, presented in
  Fig.~\ref{oh35}, is the result of combining eight independent
  observations of 
  the source obtained between 2006 and 2009 as part of the IRS calibration
  program (earlier versions of the spectrum can be found
  in \cite{far07} and \cite{arm07}).

\subsection{General characteristics of the OH spectra}
\label{sec:gen}

The OH spectra displayed in Fig.~\ref{mrk231} show a diversity of line
shapes. The ground-state OH119 and OH79 doublets (panels a and b) exhibit
prominent P-Cygni profiles, indicative of outflowing gas with the absorption
produced in front of, and the emission feature laterally adjacent to and
behind the far-IR source. 
Absorption in OH119 is found up to a blueshifted velocity of
$-1600$ \kms, while the case of OH79 is uncertain due to contamination by
\hdo. Emission in OH79 is detected up to a velocity of $\approx770$ \kms\ from
the red OH component. The third ground-state transition within the PACS range,
the OH53.3 doublet (panel e), does not show any emission feature. Its
blueshifted absorption extends up to at least $-1000$ \kms, being contaminated
by the very high-lying OH53 doublet at more negative velocities (also shown in
panel e). The Spitzer IRS OH35 spectrum (Fig.~\ref{oh35}), which has 
significantly lower spectral resolution ($500$ \kms), shows absorption that
peaks at central velocities with no emission feature;
the absorption on the blue side is more prominent than on the red side.
Detection of OH79, OH53.3, and OH35 implies that OH119, with a much higher
opacity (F10), is optically thick. However, the peak absorption in OH119 is 
only $30$\% of the continuum, indicating that the OH119 doublet at a given
velocity only covers a fraction of the total 119 $\mu$m continuum, and/or
that the 119 $\mu$m transition is very excited\footnote{The
absorption strength of the transition approaches zero if the excitation
temperature of the transition approaches the dust temperature.}.

The excited OH84 and OH65 doublets (panels c and d) do not show any emission
feature either, but display prominent blueshifted absorption. It is worth
noting that while OH65 shows absorption up to $\sim-1500$ \kms,
the less excited OH84 doublet (with lower S/N) only shows absorption up
to $\sim-1000$ \kms. The reliability of the extreme OH65 blueshifted absorption
is discussed in Sect.~\ref{sec:blue}. 

While the peak absorption in the OH119 and OH79 doublets is blueshifted
by $-300$ and $-240$ \kms, respectively, relative to the blue component of
the doublet, the OH53.3 and OH35 peak closer to central velocities. 
The increase in line excitation along the $\Pi_{3/2}$ ladder also shifts
the peak absorption toward rest velocities, with the high-lying OH65 and 
OH53 doublets peaking at nearly $v=0$ \kms. The latter transition does not
show any hint of blueshifted absorption to within the S/N. These
velocity shifts suggest that the excited lines trace an outflow region not
entirely coincident with that probed by the ground-state OH119 and OH79. The
OH119 and OH65 spectra also show blueshifted absorption at velocities 
significantly higher than the line wing emission in CO and HCN,
which is observed just out to $\sim800$ \kms\ 
\citep{fer10,cic12,aal12}. 

Along the $\Pi_{1/2}$ ladder, the high-lying OH71 doublet (panel f) also
peaks at rest velocities, with possible blueshifted absorption that is
uncertain due to the proximity of a strong \hdo\ line at $\approx-500$
\kms. Hints of emission are also seen at redshifted velocities in the OH71
doublet.

The only OH transition with emission dominating over absorption is the OH163
doublet (panel h). The features are slightly redshifted (by $\approx50$ \kms),
and there are hints of redshifted emission at velocities higher than 700 \kms\
(in line with the emission feature in OH79 at extreme, positive
velocities). The OH163 doublet also shows blueshifted absorption up
to $-350$ \kms, as well as a relatively weak emission feature at $-480$
\kms\ that is probably attributable to ($\sim100$ \kms\ redshifted) CO
(16-15).

\subsection{Potential contaminations}
\label{sec:cont}

Potential contaminations by lines of species other than OH are indicated in
Fig.~\ref{mrk231}, and Fig.~\ref{oh84} compares the OH65 and OH84 spectra of
Mrk~231, Arp~220, and NGC~4418. In OH119, the redshift component
$J=7/2^+-5/2^-$ of the excited $N=3-2$ CH doublet
($E_{\mathrm{lower}}\approx105$ K, $118.705$ $\mu$m) is not expected to
contaminate the blueshifted OH absorption at 
$-1300$ \kms, because the blue component $J=7/2^--5/2^+$ of the doublet is
undetected. Likewise, there are no apparent features at the positions of the
two indicated p-H$_2$O$^+$ lines ($4_{14}-3_{03}\,7/2-5/2$ and
$9/2-7/2$), also not detected in Arp~220 \citep[][hereafter
G-A13]{gon13}. However, the CH$^+$ $3-2$ line could be contaminating the
$^{18}$OH doublet (F10), as the ground 
transition of CH$^+$ is bright in Mrk~231 \citep{wer10}.  

In the OH79 profile, the \hdo\ \t423312\ line at $-1400$ \kms\
($78.742$ $\mu$m, $E_{\mathrm{lower}}\approx250$ K) is clearly detected
\citep[F10,][hereafter G-A10]{gon10}, though the prominent blueshifted
absorption apparent in the GT observation is not confirmed with the OT2
observation (Table~\ref{tab:obs}).  
On the other hand, the higher-lying \hdo\ \t615524\ line at $-720$ \kms\
($78.928$ $\mu$m, $E_{\mathrm{lower}}\approx600$ K) was not seen in the GT
observation, but a spectral feature in the OT2 observation makes the case
uncertain. 

In the OH84 profile, 
the only potential contamination is due to NH$_3$ (Fig.~\ref{oh84}a), 
which is expected to generate an absorption wing-like feature between $300$
and $1300$ \kms\ associated with the $(6,K)a-(5,K)s)$ lines (G-A12). However,
from the uncontaminated $(6,K)s-(5,K)a)$ NH$_3$ lines at $83.4-83.9$ $\mu$m,
only the $(6,5)s-(5,5)a)$ line is (marginally) detected, thus no significant
contamination by NH$_3$ to OH84 is expected.

In the OH65 profile, no spectral feature coinciding with the p-H$_2$O$^+$
$3_{30}-2_{21}\,5/2-3/2$ line at $-1000$ \kms\ is found (also undetected
in Arp~220, Fig.~\ref{oh84}b). The \hdo\ \t625514\ line at $\approx160$
\kms\ could contaminate the main OH feature (G-A12), and o-H$_2$O$^+$
$3_{31}-2_{20}\,7/2-5/2$ and NH$_2$ \t422313\ are probably contributing to the
$^{18}$OH absorption feature. 

A weak contribution by the \hdo\ \t533422\ line 
is expected around $-700$ \kms\ in the OH53.3 profile
(Fig.~\ref{mrk231}e). 
The OH71 spectral region is complex (Fig.~\ref{mrk231}f), with
possible baseline curvature and
contaminations on the blue side by \hdo\ \t524413\ at $-430$ \kms\ 
($E_{\mathrm{lower}}\approx400$ K, $71.067$ $\mu$m) and possibly H$_3$O$^+$
$4_1^--3_1^+$ and $4_0^--3_0^+$ (both detected in NGC~4418 and Arp~220,
G-A13). These lines are expected shortward of an apparent absorption
pedestal $\gtrsim1000$ \kms\ wide, 
to which both \hdo\ and OH may be contributing. On the
red side, absorption by the very high-lying \hdo\ \t717606\ line at
$1550$ \kms\ is detected. 

\begin{figure}
   \centering
   \includegraphics[width=8.5cm]{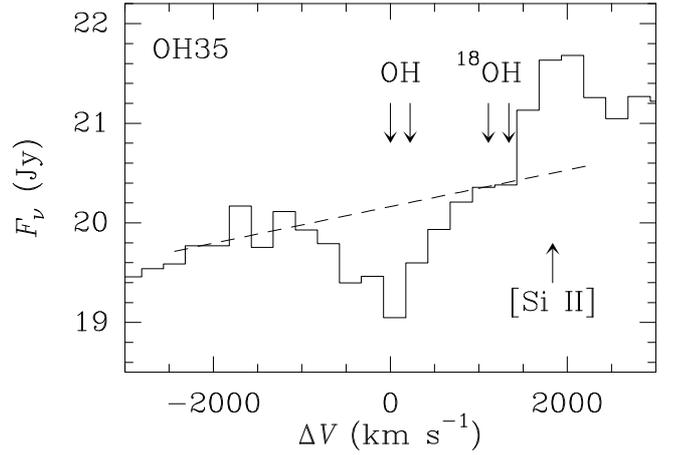}
   \caption{Spitzer IRS long-high (LH) spectrum around the OH35 doublet,
     with the adopted baseline (dashed line). The emission feature at
     $\sim2000$ \kms\ is the [Si II] line at 34.815 $\mu$m. The spectral
     resolution is 500 \kms. 
    } 
    \label{oh35}
    \end{figure}

\begin{figure}
   \centering
   \includegraphics[width=8.5cm]{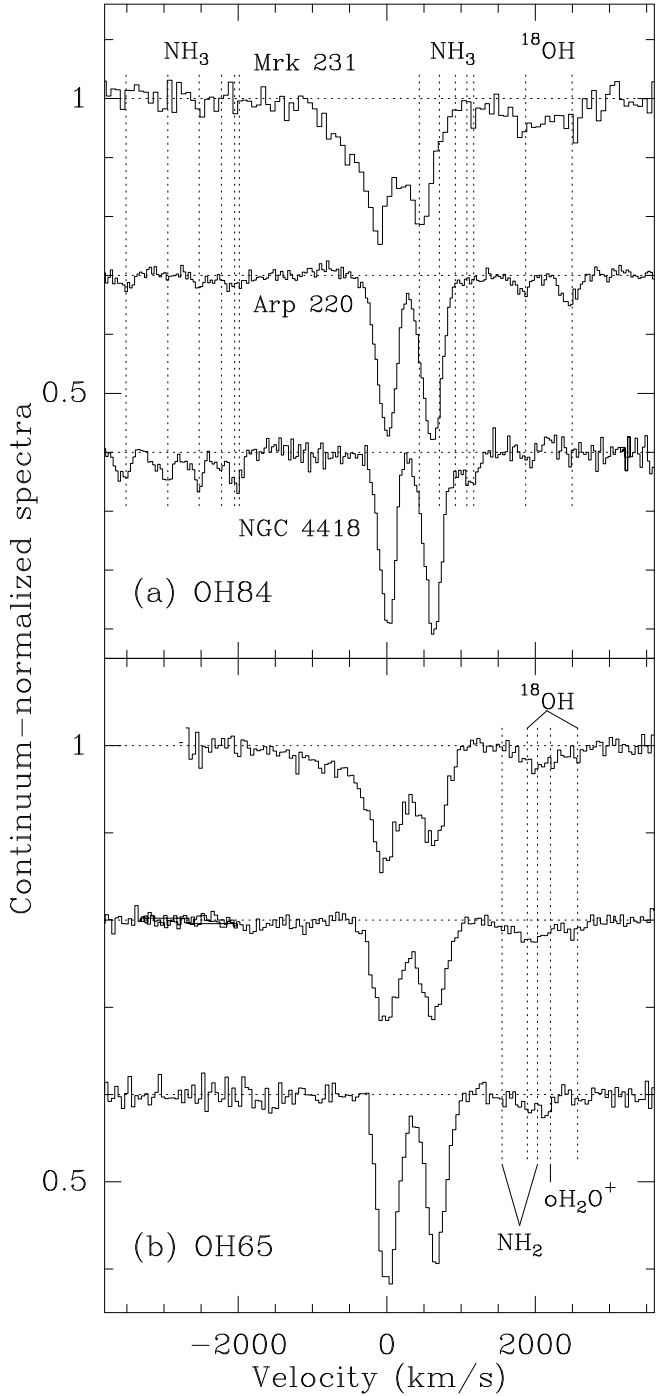}
   \caption{Spectra around {\bf (a)} the OH84 doublet, and {\bf (b)} the OH65
     doublet, in Mrk~231 ({\it upper spectra}), Arp~220 ({\it middle}), and
     NGC~4418 ({\it lower}). The vertical-dotted lines in {\bf (a)}
     indicate the positions of NH$_3$ and $^{18}$OH lines. The NH$_3$ lines at
     $v<-2000$ \kms\ are relatively strong in NGC~4418, indicating
     contribution by NH$_3$ to the absorption at $v\sim1000$ \kms. This is not
     the case of Mrk~231, so that there is no significant contamination at
     $v\sim1000$ \kms. The $^{18}$OH66 doublet in {\bf (b)} is complex with
     probable contributions by o-H$_2$O$^+$ and NH$_2$.
    } 
    \label{oh84}
    \end{figure}

The OH56 spectrum shows strong absorption by the \hdo\ \t431322\ transition at
$2300$ \kms.
The OH163 spectrum is free of contaminations, except for the relatively
weak CO (16-15) emission line.  
Finally, the OH35 spectrum (Fig.~\ref{oh35}) shows the [Si II] line in
  emission at $\sim2000$ \kms, precluding the measurement of the
  $^{18}$OH35 doublet profile.

\subsection{Blueshifted line wings and the OH65 spectrum}
\label{sec:blue}

As mentioned in Sect.~\ref{sec:gen}, at least the OH119, OH79, OH53.3, OH84,
and OH65 doublets show high-velocity absorption wings extending up
to at least $-1000$ \kms\ from the blue component of the doublets; the possible
OH71 line wing is contaminated by \hdo. 

   \begin{figure}
   \centering
   \includegraphics[width=8.8cm]{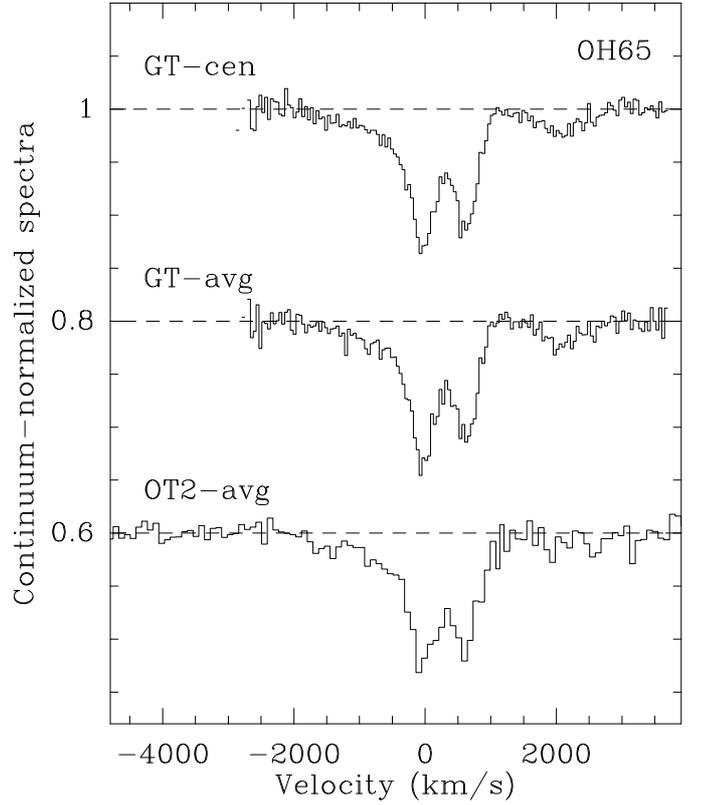}
   \caption{OH65 doublet toward Mrk 231. Three spectra are compared:
     GT-cen is the central spaxel of the GT observations, GT-avg is the
     average spectrum of the three brightest spaxels in the GT observations,
     and OT2-avg denotes the average spectrum of the three brightest spaxels
     in the OT2 observations. All three spectra show a prominent blueshifted
     line wing extending up to at least $\sim-1500$ \kms.}
    \label{oh65}
    \end{figure}

While the OH79 and OH53.3 are contaminated at velocities more blueshifted than
$-1000$ \kms, the OH119, OH84, and OH65 doublets along the $\Pi_{3/2}$ ladder
are uncontaminated throughout the blue range and probe the outflowing gas at
the most extreme velocities. OH119 shows absorption up to $-1600$ \kms, but
OH84 only up to $-1000$ \kms\ with a lower S/N. Very intriguing then is
the apparent strong absorption in the more excited OH65 doublet, up to
velocities of at least $\sim-1500$ \kms. Figure~\ref{oh65} compares three
spectra of the OH65 doublet. 
Both the GT spectra of the central spaxel (GT-cen, see
Table~\ref{tab:obs}) and of the three brightest spaxels (GT-avg, also shown in
Fig.~\ref{mrk231}d) show the line wing covering a similar velocity range, 
indicating that it cannot be ascribed to pointing effects that would generate
fluctuations of the continuum level. Furthermore, the OT2-avg spectrum,
with a flat baseline (albeit with a lower S/N, though mostly at
positive velocities), also shows the line wing with a similar velocity
extent. The close agreement between the OH65 GT and OT2 high-velocity
  wing profiles together with the smaller extent of the OH84 profile provide
  strong evidence for the presence of highly excited gas at extreme
  velocities.

   \begin{figure}
   \centering
   \includegraphics[width=8cm]{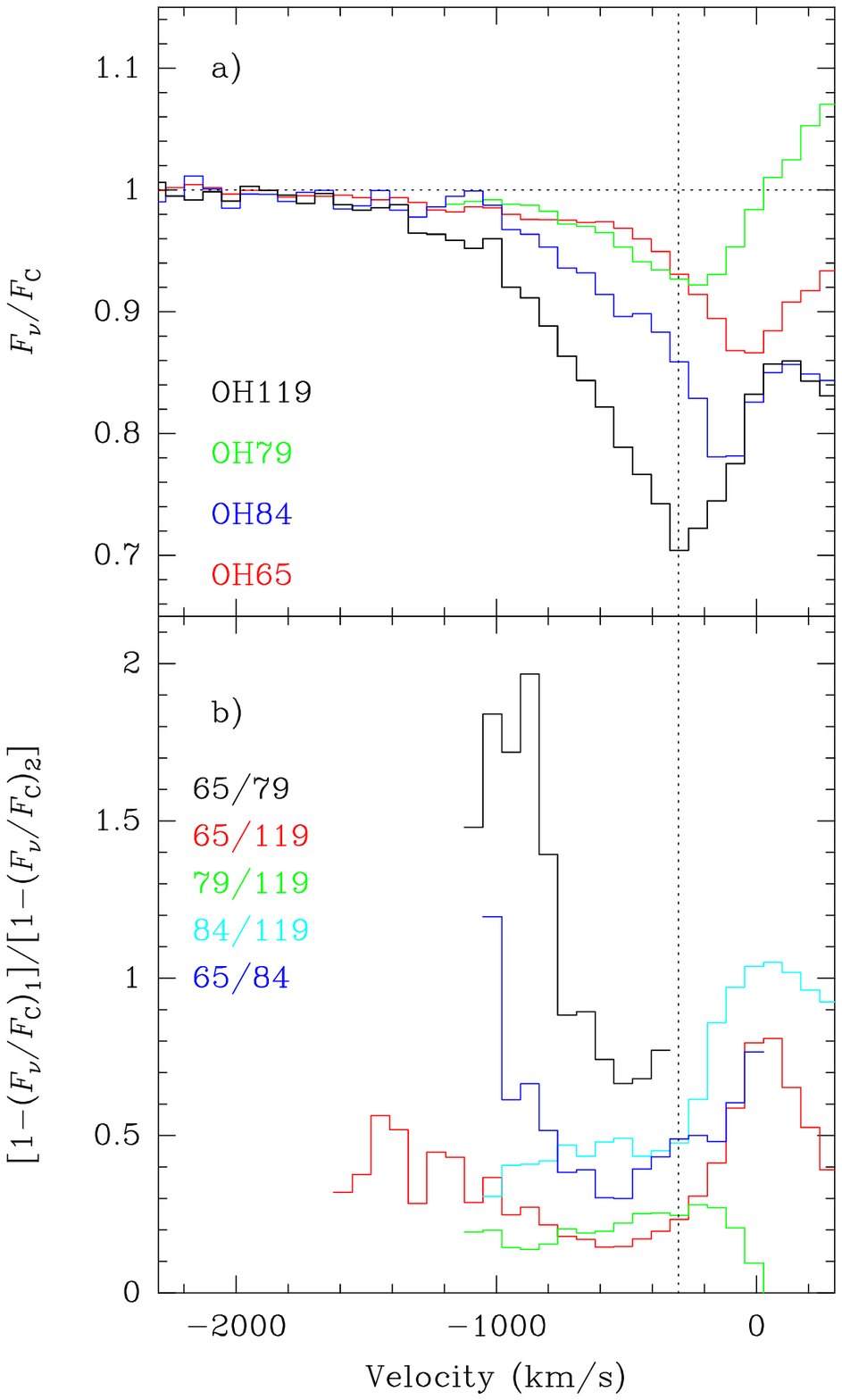}
   \caption{{\bf a)} Blueshifted line wing as observed in four OH
     doublets at 119 (black histogram), 79 (green), 65 (red), and 84 $\mu$m
     (blue). The spectra are resampled to the velocity resolution of the
     OH119 spectrum. The velocity scale is relative to the blueshifted
     component of each doublet. {\bf b)} Ratios of the absorption 
strengths $(1-F_{\nu}/F_C)$. The vertical dotted line at $-300$
\kms\ marks the limit of the outflow region for the full set of lines.}  
    \label{excit}
    \end{figure}

Figure~\ref{excit} compares the blue line wings observed in OH119, OH79,
OH84, and OH65 in more detail and shows the ratios of their absorption 
strengths. For velocities in the range $-300$ to $-1000$ \kms, the intensity
in all spectra decreases smoothly, and the line ratios are nearly constant
with the exception of the $65/79$, which increases significantly with
increasing velocity shift. At higher velocity shifts, the $65/84$ and $65/119$
ratios jump to higher values. The limited S/N of the OH84 spectrum and the
baseline uncertainties do not allow us to establish an accurate limit for the
$65/84$ ratio, but it is most likely $>0.5$.

\subsection{P-Cygni profiles and the role of far-IR extinction and geometry}
\label{sec:pcyg}

P-Cygni profiles are observed in at least the OH119 and OH79 doublets. The
upper level of the OH119 transition is only efficiently populated
from the ground-state through absorption 
of a 119 $\mu$m photon or through a collisional event, that is, there is
no efficient excitation path that involves radiative pumping to a higher-lying
level and cascade down to the $\Pi_{3/2}\, J=5/2$ level\footnote{Pumping
  the $\Pi_{1/2}\, J=3/2,5/2$ levels through absorption of $53.3$ and $35$
  $\mu$m photons and subsequent decay via $\Pi_{1/2}-\Pi_{3/2}\,
  J=3/2,5/2-5/2$ is less efficient as the latter cross-ladder
  transitions are relatively weak, but can still boost the emission
    feature relative to the absorption one by $\sim35$\%.}. Under certain 
idealized conditions, this translates into a relationship between the fluxes
of the absorption and emission features. 
Assuming spherical symmetry, a two-level system, pure radiative
excitation (i.e. negligible collisional excitation), and an envelope size that
is much larger than the size of the continuum source (i.e. with negligible
extinction of line photons emitted from behind the continuum source), then
statistical equilibrium of the populations and complete redistribution in
angles ensures equal number of absorption and emission events as seen by an
observer that does not spatially resolve the outflow. 
In that case, and due to conservation of continuum
photons, the outflowing gas would have the overall effect of redistributing
the continuum photons in velocity space, generating a redshifted emission
feature as strong as the blueshifted absorption feature\footnote{In
spherical symmetry, all observers located at the same distance from
the source would detect exactly the same spectrum, and since we assume that
there is neither cooling in the line (no collisions) nor absorption of
line-emitted photons by dust, conservation of the continuum emission
holds regardless of the line opacity, implying equal rates of
absorption and emission events.}  
\citep[this is analogous to
the \hdo\ 6 $\mu$m band in Orion BN/KL where the P-branch is observed in
emission and the R-branch in absorption,][]{gon98}. In Mrk~231, however, the
emission feature in the OH119 doublet is five
times weaker than the absorption feature, revealing $(i)$ the importance of
extinction of line-emitted photons arising from the back side of the
far-IR source, and/or $(ii)$ significant departures from spherical symmetry,
with the outflow mainly directed toward the observer
(e.g. bipolar) and/or the receding component intrinsically less prominent
than the approaching one. These conclusions are strengthened if collisions
in a warm and dense region, suggested by the bright emission observed in
the ground-state transitions of several species \citep{wer10}, enhance the
OH119 excitation in the receding emitting gas. 

   \begin{figure}
   \centering
   \includegraphics[width=8.8cm]{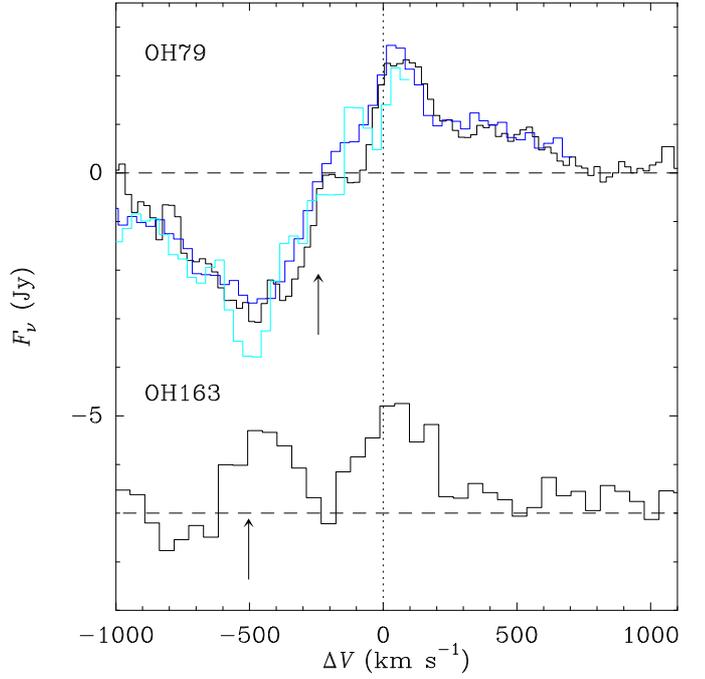}
   \caption{Comparison between the OH79 and OH163 (vertically shifted for
     clarity) continuum-subtracted spectra. The velocity scale is here
relative to the red $\Lambda-$component of each doublet, with the aim of
directly comparing the emission features. The vertical arrows indicate
the positions of the blue components of the doublets.}  
    \label{oh79_163}
    \end{figure}

Since the OH84 doublet does not show a redshifted emission feature
  (Fig.~\ref{mrk231}c and \ref{oh84}a), we argue in Sect.~\ref{sec:colhvc} that
  the circumnuclear outflowing gas does not significantly contribute to the
  OH119 emission feature (at $v>200$ \kms\ from the red component, and for the
  reasons pointed out above),  
  which then arises primarily from a spatially extended 
  low-excitation component where the far-IR extinction is presumably low
  (Sect.~\ref{sec:lec}). Assuming that collisional excitation of OH in this
  component is negligible (i.e. scattering of dust-emitted photons at 119
  $\mu$m is responsible for the observed emission), the strength of the OH119
  emission feature relative to the observed continuum level then measures,
  under some conditions, the clumpiness of the receding gas. If the receding
  OH were intercepting the full 119 $\mu$m continuum emitted toward the back
  $2\pi$ sr within a velocity interval $\Delta v$, and neglecting again
  extinction, one would expect an emission feature as strong as $\sim50$\% of
  $F_{119}\Delta v$ ($F_{119}$ is the observed continuum flux density at 119
  $\mu$m), i.e. $\sim6\times10^3$ Jy \kms\ for $\Delta v\sim500$ \kms\ (see
  Fig.~\ref{mrk231}a)\footnote{It is also assumed that the red component
of the doublet dominates the emission feature and that this emission is 
radiatively decoupled from absorbing foreground gas; the redshifted
emission due to the blue component of the doublet is cancelled or
  blocked by the blueshifted absorption due to the red doublet
component. Note also that we use the observed line width of the emission
feature as an (approximate) proxy for the velocity range within which
$\tau_{\mathrm{OH}119}\ge 1$.}.  
  The observed emission    
  feature, however, accounts for only $\sim1.2\times10^3$ Jy \kms, thus
  suggesting a covering factor of $\sim20$\% for the above conditions. If the
  receding OH is behind the source of 119 $\mu$m continuum emission, only
  $\sim20$\% of that continuum within $\Delta v=500$ \kms\ is thus intercepted
  and reemitted by the OH, indicating high clumpiness of the spatially
  extended reemitting OH. This is consistent with the clumpiness of the
  outflow inferred by \cite{aal12} from HCN emission. 

The ground-state OH79 doublet shows a relatively strong emission feature, with
a flux that is nearly $75$\% of the absorption flux. Since extinction at 79
$\mu$m is higher than at 119 $\mu$m, the prominent OH79 emission feature
indicates the effect of radiative pumping through absorption of 53.3 and 35
$\mu$m continuum photons, and subsequent cascading down to the ground
state via the 163 and 79 $\mu$m transitions (G-A12). The
OH163 doublet (panel h) is indeed mostly observed in emission, qualitatively
matching the pumping scheme. More quantitatively, Fig.~\ref{oh79_163}
  compares in detail the OH79 and OH163 profiles, with the velocity scale
  relative to the red component of each doublet. The flux of the OH163
doublet between 0 and 210 \kms\ is $\approx400$ Jy \kms, about 10\%
higher than the flux emitted in the OH79 doublet in the same velocity 
interval (i.e. the narrow-emission feature, 365 Jy \kms). Owing to the
contribution to OH79 by a prominent redshifted line wing that is weak in
  OH163, the intrinsic flux of the narrow OH79 emission feature without the
wing contribution is estimated to be $\sim200$ Jy \kms, a
factor of $\sim2$ weaker than that of the OH163. Since every 163
$\mu$m line-emitted photon should be accompanied by a 79 $\mu$m one
(Fig.~\ref{diag}), but the OH79 emission is additionally boosted by
  direct scattering of 79 $\mu$m dust-emitted photons (which does not involve
  emission in the OH163 doublet), the difference in fluxes\footnote{
      Fluxes in units of Jy km s$^{-1}$ are proportional to the rate of
      de-excitation events in the line.} indicates,
regardless of geometry, that indeed significant extinction affects the narrow
emission feature in the OH79 doublet.  

The prominent redshifted line-wing observed in emission in the OH79
doublet is also remarkable, with a flux of $\approx400$ Jy km/s between 
  210 and 800 km/s (Fig.~\ref{oh79_163}). Some hints of emission are also
  found in the OH163 doublet at $>210$ \kms, with an uncertain ($\pm50$\%)
  flux of $\sim185$ Jy \kms. This flux is weaker than that measured in
  the OH79 redshifted wing, thus indicating that the emission in this OH79
  wing is not significantly extincted and that resonant scattering of 79
  $\mu$m dust-emitted photons probably dominates the OH79 wing-emission
  feature. It is thus plausible that the OH79 redshifted wing is
  more extended than the source of the far-IR emission. The velocity
    extent of the OH79 redshifted emission feature is
    similar to that of other molecular lines at millimeter wavelengths
    \citep{fer10,cic12,aal12}, suggesting a similar spatial origin.

The ground-state OH53.3 doublet (Fig.~\ref{mrk231}e)
does not show an emission feature, due in part to the slightly higher
  chance for an OH molecule in the $\Pi_{1/2} \,\, J=3/2$ level to decay via
  the 163 $\mu$m transition instead of directly emitting a $53.3$ $\mu$m
  photon (Fig.~\ref{diag}), but also further indicating the role of
extinction. Similarly, the OH35 doublet profile (Fig.~\ref{oh35}) only
  shows absorption, as expected given that essentially
all molecules pumped to the upper $\Pi_{1/2} \,\, J=5/2$ level will relax by
emitting a 99 $\mu$m photon along the $\Pi_{1/2}$ ladder (instead of
emitting a 35 $\mu$m photon). There is significant redshifted absorption in
OH35, indicating that there is still 35 $\mu$m continuum emission behind part
of the receding gas.

It is then intriguing that the high-lying OH71 doublet
(Fig.~\ref{mrk231}f) shows hints of redshifted emission, with a flux of 150
Jy \kms. The reliability of this emission feature is uncertain,
however, as it shows different strengths in the central-spaxel and averaged
spectra. The line should be formed very close to a warm source of far-IR
radiation, which is probably optically thick at these wavelengths. If the
feature is not an artifact of the baseline, inhomogeneities of the dust
extinction in the nuclear region and geometry effects would be required to
account for it.

\subsection{Absorption at central velocities: a very excited, quiescent
gas component} 
\label{sec:cenv}

In Fig.~\ref{mrk231}, the excited OH lines (other than OH163) show strong
absorption at central velocities, similar to NGC~4418 and Arp~220
(G-A12). This reveals the presence of a very excited, non-outflowing
component in the nuclear region of Mrk~231. However, no trace of a relatively
narrow absorption feature is found at central velocities in the OH119 and OH79
doublets. This is conspicuous, because 
extinction at 119/79 $\mu$m would strengthen
  the line absorption relative to the emission feature, as argued in
  Sect.~\ref{sec:pcyg} for the outflowing gas. In a quiescent component, the
  absorption and reemission occur at the same central velocities, so that one
  would expect a resulting central absorption 
  feature in spherical symmetry. The other ground-state transitions, the
  OH53.3 and OH35 doublets, do show strong absorption at central velocities. 

The lack of measurable absorption at central velocities in the OH119
transition may be partially due to the fact that the enclosed dust is
very warm ($>100$ K) and compact, thus emitting weakly at 119 $\mu$m 
  compared with the total emission at this wavelength. In this case, the
OH119 absorption would be strongly diluted within the observed 119
$\mu$m continuum emission, whose main contribution would arise from
more extended regions devoid of quiescent OH. In addition, collisional
excitation in a warm and dense circumnuclear component would also excite the
OH molecules to the level of near radiative equilibrium with the dust,
thus producing negligible absorption. A potentially important effect is also
resonant scattering of dust-emitted photons in the OH119 doublet in a
flattened structure (e.g. a torus or disk) seen nearly face-on or
moderately inclined, which would tend to cancel the absorption produced
toward the strongest continuum source. It is worth noting that since
the OH119 transition is ground-state, this
process could work on spatial scales significantly larger than the region
responsible for the high-excitation absorption observed at systemic
velocities (see also Sect.~\ref{sec:qc}).

While in the OH79 transition the pumping via the 
$53.3$ and $35$ $\mu$m transitions enhances the reemission, in the OH53.3
doublet the upper level is 
higher in energy and hence more difficult to excite collisionally, and
reemission is less favored because of the competing de-excitation path
via the OH163 transition (Fig.~\ref{diag}).  

The OH119 and OH79 profiles of Mrk~231 are in this respect very
different from those observed in Arp~220, which shows in these doublets
strong absorption at central velocities (G-A12). The OH spectra of the
high-lying lines are more similar at central velocities
(Fig.~\ref{oh84}), indicating that both sources have highly excited
OH. This indicates that the components that are responsible for the
ground-state absorption in Arp~220 at central velocities, that is,
\chalo\ and \cext\ (G-A12), are absent in Mrk~231, which is consistent 
with the face-on view of the disk at kpc scales.

\subsection{High OH optical depths}
\label{sec:ohopacity}

The equivalent widths of the OH35 and OH53.3 doublets are 
$\approx45$ and $\approx120$ \kms, respectively. For optically thin
absorption, ignoring reemission in the lines and assuming that the OH
molecules are covering the whole continuum source at the corresponding
wavelengths, the equivalent width of a doublet (in units of velocity) is given
by  
\begin{equation}
W_{\mathrm{eq}}=\lambda^3 g_u A_{ul} \frac{N_{\mathrm{OH,gr}}}{8\pi g_l},
\label{eq:weq}
\end{equation}
where $\lambda$ is the wavelength, $A_{ul}$ is the Einstein coefficient for
spontaneous emission, $g_u$ ($g_l$) is the degeneracy of the upper (lower)
level, and $N_{\mathrm{OH,gr}}$ is the OH column density in the two
lambda-doubling states of the ground $\Pi_{3/2} \, J=3/2$ rotational
level. Therefore, in the optically thin limit, the OH53.3-to-OH35 equivalent
width ratio is
$W_{\mathrm{eq}}(\mathrm{OH53.3})/W_{\mathrm{eq}}(\mathrm{OH35})\approx6.4$, 
while the observed ratio is $\sim3$. This indicates that opacity effects are
important even in the OH53.3 doublet, which is less optically thick than the
OH79 doublet. 

Using the OH35 doublet, the lowest optical depth ground-state doublet,
in eq.~(\ref{eq:weq}) gives $N_{\mathrm{OH,gr}}\approx1.1\times10^{17}$ \cmd,
which is a lower limit for $N_{\mathrm{OH}}$ because $(i)$ a significant
fraction of molecules is in excited levels; $(ii)$ extinction at 35 $\mu$m
only enables the detection of OH in the external layers of the continuum
source; and $(iii)$ the OH may not be covering the whole 35 $\mu$m continuum
source.

\subsection{$^{18}$OH}
\label{sec:18ohini}

Up to three $^{18}$OH doublets are detected within the PACS range, at
120, 85, and 66 $\mu$m. While the $^{18}$OH120 doublet may be partially
contaminated by CH$^+$, and $^{18}$OH66 by NH$_2$ and H$_2$O$^+$, the
$^{18}$OH85 doublet is 
free from contamination, with an integrated flux about five times weaker
than the OH84 doublet. This confirms the strong enhancement of $^{18}$OH in
Mrk~231 (F10). It is also worth noting that while the OH84 profile
shows a dip in absorption between the two lambda-doubling components, a nearly
continuous bridge of absorption is seen between the $^{18}$OH components
(probable contamination makes the case uncertain in $^{18}$OH65, where
the absorption peaks in between the two lambda components). This may suggest a
relative enhancement of $^{18}$OH in the outflowing gas.

\section{Models and interpretation}
\label{sec:analysis}

\subsection{Radiative-transfer models}
\label{sec:rt}

\subsubsection{Overview}
\label{sec:ovw}

To estimate the physical properties of the molecular outflow as
derived from the OH doublets, we analyzed the OH line
profiles and fluxes using radiative-transfer models. We used the
code described in \cite{gon99}, which calculates in spherical
symmetry the line excitation due to the dust emission and collisions with
H$_2$, and includes opacity effects (i.e. radiative trapping), non-local
effects, velocity gradients, extinction by dust, and line overlaps
\citep{gon97}. For a given model, the code first calculates the 
statistical-equilibrium populations in all shells that make up the
source, and then the emerging line shapes are computed, convolved with
the PACS spectral resolution, and compared directly with the
observations.  

   \begin{figure}
   \centering
   \includegraphics[width=8.8cm]{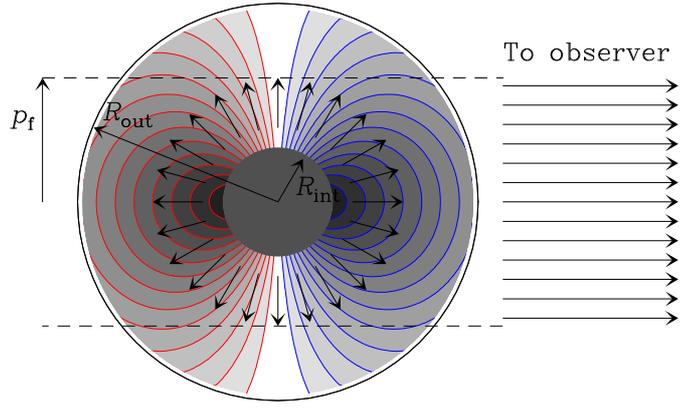}
   \caption{Schematic representation of a given source component. It consists
     of $(i)$ a central source of far-IR emission characterized by its radius
     ($R_{\mathrm{int}}$), dust temperature ($T_{\mathrm{dust}}$), and optical
     depth at 100 $\mu$m along a radial path ($\tau_{100}$), and $(ii)$ the
     surrounding (outflowing) gas, with external radius
     $R_{\mathrm{out}}$, which is mixed with dust. The gas expands radially
     with a velocity field $v(r)$ and H nuclei density $n(r)$, such that
     $n_{\mathrm{OH}}\times r^2\times v$ is constant (i.e. the mass-outflow
     rate is constant). The dust in the outflowing envelope has
     $\tau_{50}=0.5$ along a radial path, and $T_{\mathrm{dust}}\sim r^{-0.4}$.
     In some models, departures from spherical symmetry are simulated by
     calculating the emerging fluxes only for impact parameters lower than
     $p_{\mathrm{f}}$ (i.e. between the two dashed lines). The blue and red
curves (and grayscale) show the isocontours of line-of-sight velocities
     (blue: approaching; red: receding) for a decelerating outflow; the
     darkest colors correspond to the highest (approaching or receding)
     velocities. In our best-fit models for the {\em excited} OH lines
     (the HVC component in Sect.~\ref{sec:comp}), the
     outflowing envelope is less extended than in this representation
     ($R_{\mathrm{out}}/R_{\mathrm{int}}\lesssim1.5$) and is collimated
     ($p_{\mathrm{f}}< R_{\mathrm{out}}$).  
    } 
    \label{outflow}
    \end{figure}

   \begin{table*}
      \caption[]{Parameters for the modeling of the OH outflow.}
         \label{tab:par}
          \begin{tabular}{llll}   
            \hline
            \noalign{\smallskip}
Parameter  & Units & Meaning &  Explored range (HVC) \\  
            \noalign{\smallskip}
            \hline
            \noalign{\smallskip}
$R_{\mathrm{int}}$ & pc & Radius of the far-IR continuum source$^{\mathrm{a}}$
&  $^{\mathrm{e}}$ \\ 
$T_{\mathrm{dust}}$ & K & Dust temperature of the far-IR continuum source &
$90-200$ \\
$\tau_{100}$ &  & Continuum optical depth at 100 $\mu$m along a radial
($R_{\mathrm{int}}$) path &  $0.5-4$ \\
$R_{\mathrm{out}}/R_{\mathrm{int}}$ & & Radius of the outflowing envelope
  relative to $R_{\mathrm{int}}$  &  $1.1-2.5$$^{\mathrm{f}}$ \\
$v_{\mathrm{int}}$ & \kms & Gas velocity at
$R_{\mathrm{int}}$$^{\mathrm{b,c}}$  & $1300-1700$$^{\mathrm{g}}$ \\
$v_{\mathrm{out}}$ & \kms & Gas velocity at
$R_{\mathrm{out}}$$^{\mathrm{b,c}}$ &  $100-400$$^{\mathrm{g}}$ \\ 
$N_{\mathrm{OH}}$ & cm$^{-2}$ & OH column density from
$R_{\mathrm{int}}$ to $R_{\mathrm{out}}$$^{\mathrm{b,c}}$ &
$(0.5-5)\times10^{17}$  \\   
$p_{\mathrm{f}}$ & pc & Limiting impact parameter for the calculation of
emerging fluxes  & $R_{\mathrm{int}}-R_{\mathrm{out}}$$^{\mathrm{h}}$ \\
$f$            &   &  Scaling factor $^{\mathrm{d}}$ &  \\
            \noalign{\smallskip}
            \hline
         \end{tabular} 
\begin{list}{}{}
\item[$^{\mathrm{a}}$] It coincides with the inner radius of the OH envelope.
\item[$^{\mathrm{b}}$] A uniform velocity gradient is adopted, so that the
  velocity field is given by $v(r)=v_{\mathrm{int}}+dv/dr \, (r-R_{\mathrm{int}})$.
\item[$^{\mathrm{c}}$] A constant mass-outflow rate is adopted, so that
  $n_{\mathrm{OH}}\times r^2\times v$ is independent of $r$.
\item[$^{\mathrm{d}}$] Representing either partial coverage by OH of the
  continuum source (a clumpy outflow, $f<1$), or an ensemble of independent
  sources ($f>1$). $\dot{M}$ scales as $\sqrt{f}$. $f$ is not a fitting
  parameter, but indicates that the modeled source size is
  effective. Nevertheless, we argue in Sect.~\ref{sec:discussion} that
  $f\sim1$ for the QC, and in Sect.~\ref{sec:colhvc} that $f\gtrsim0.45$ for
  the HVC. 
\item[$^{\mathrm{e}}$] For a given model, $R_{\mathrm{int}}$ is fixed to
    give the correct absolute fluxes.
\item[$^{\mathrm{f}}$] See Fig.~\ref{hvc}.
\item[$^{\mathrm{g}}$] Accelerating velocity fields have been tried as
well, but they yield poor fits to both the line profiles and the flux ratios.
\item[$^{\mathrm{h}}$] $p_{\mathrm{f}}=R_{\mathrm{out}}$ in
    spherical symmetry, while $p_{\mathrm{f}}=R_{\mathrm{int}}$ simulates an
    outflow directed mainly toward the observer.
\end{list}
   \end{table*}

As shown below, at least three components (two outflowing components with
  different velocity fields, spatial extents, and far-IR radiation sources,
  and one relatively quiescent component with little or no outflowing motion,
  hereafter referred to as the QC) 
are required to obtained a reasonable match to the observed line profiles. The
different components are modeled separately, and the corresponding 
  emerging flux densities are then summed up together (i.e. it is assumed
that the different 
components do not simultaneously overlap along the line of sight and in the
projected velocity). Figure~\ref{outflow} depicts the generic model for a given
outflowing component (the QC component described
  below is modeled as in G-A12). 
A central source of far-IR radiation is characterized by its 
radius $R_{\mathrm{int}}$, dust temperature $T_{\mathrm{dust}}$, and optical
depth at 100 $\mu$m along a radial path $\tau_{100}$. This is surrounded by an
envelope of outflowing molecular gas with external radius $R_{\mathrm{out}}$. 
The OH is mixed with the dust in the envelope, where $T_{\mathrm{dust}}\sim
r^{-0.4}$ \citep[e.g.][]{ada91} and $\tau_{50}=0.5$ between
$R_{\mathrm{int}}$ and $R_{\mathrm{out}}$\footnote{This only applies to
    the high-velocity component (HVC) discussed below, where a column of
    $N_{\mathrm{H}}\sim10^{23}$ \cmd\ for the outflowing shell is estimated
    (Table~\ref{tab:ener}). A value of $\tau_{50}\approx0.25$ across the
    outflowing gas is then expected (for $N_{\mathrm{H}}\sim4\times10^{23}$
    \cmd\ per unit of $\tau_{50}$, G-A12). We doubled that number to
    roughly 
    simulate illumination by an external radiation field and/or emitting
    clumps mixed with the outflowing gas, though this has a weak effect
    on results because the excitation is dominated by the central far-IR
    source.}.  
The gas is outflowing in spherical symmetry with 
velocity and H$_2$ density profiles $v(r)$ and $n(r)$, respectively. To
decrease the number of free parameters, we imposed a constant velocity 
gradient (i.e. $v(r)=v_{\mathrm{int}}+dv/dr \, (r-R_{\mathrm{int}})$, where
$dv/dr=(v_{\mathrm{out}}-v_{\mathrm{int}})/(R_{\mathrm{out}}-R_{\mathrm{int}})$
is constant) and a constant mass-outflow rate (mass conservation then
implies that $n_{\mathrm{OH}}\times r^2\times v$ is independent of $r$). 

A constant OH abundance relative to H nuclei, 
$X_{\mathrm{OH}}=2.5\times10^{-6}$, was adopted (S11), 
as derived to within a factor of $\sim2$ in the nuclear regions of
NGC~4418 and Arp~220 (G-A12). This value is consistent with models of XDRs and
CRDRs with relatively high ionization rates (e.g. Meijerink et al. 2011), 
  that is, in the circumnuclear region of Mrk~231. In more extended
regions (i.e. in the low-excitation component discussed in
Sect.~\ref{sec:lec}), the OH abundance may decrease 
depending on the reaction of OH with other species and photodissociation
processes, or its ability to freeze-out as the outflow expands. 
We retain below the dependence of our mass
and energy estimates on $X_{\mathrm{OH}}$ so that our results can be easily
rescaled.

According to the results shown below, strict spherical symmetry is not 
an accurate approach in some models, and gas outflowing along two
approaching and receding cocoons (i.e. with little gas expanding along
the plane of sky) is favored. This is roughly simulated by including the free
parameter $p_{\mathrm{f}}$, such that the emerging fluxes are calculated
only for impact parameters $p<p_{\mathrm{f}}$ (i.e. for rays within the cylinder
depicted with dashed lines in Fig.~\ref{outflow})\footnote{Note that this is
  only an approximation, as the level populations are calculated in spherical
  symmetry.}. Finally, the continuum-subtracted emerging profiles of a
given component can be multiplied by a factor $f\lessgtr1$, which represents
either partial OH covering of the far-IR source (i.e. a clumpy outflow,
$f<1$), or an ensemble of independent outflows ($f>1$, see below).

The free parameters for each component are then $R_{\mathrm{int}}$,
$T_{\mathrm{dust}}$, $\tau_{100}$, $R_{\mathrm{out}}/R_{\mathrm{int}}$, 
$v_{\mathrm{int}}$, $v_{\mathrm{out}}$, $N_{\mathrm{OH}}$,
$p_{\mathrm{f}}/R_{\mathrm{out}}$, and $f$, and are listed in Table~\ref{tab:par}. 
The data that constrain the fit are the line profiles and fluxes of the
nine OH doublets. The line ratios essentially depend on
$T_{\mathrm{dust}}$, $N_{\mathrm{OH}}$, and
$R_{\mathrm{out}}/R_{\mathrm{int}}$, while the absolute fluxes also depend on
$f\,R_{\mathrm{int}}^2$. The radial column density of H nuclei in a given
component is
\begin{equation}
N_{\mathrm{H}} = X_{\mathrm{OH}}^{-1}  \int_{R_{int}}^{R_{out}} 
n_{\mathrm{OH}}(r) \, dr.
\label{eq:col}
\end{equation}
The mass-outflow rate per unit of solid angle is
\begin{equation}
\frac{d\dot{M}}{d\Omega} =  f \, m_{\mathrm{H}} \, X_{\mathrm{OH}}^{-1}
\, n_{\mathrm{OH}}(R_{\mathrm{int}}) \, R_{\mathrm{int}}^2  \, v_{\mathrm{int}}, 
\label{eq:mdot}
\end{equation}
and the total mass-outflow rate is
\begin{equation}
\dot{M} = 4 \pi \, g(p_{\mathrm{f}}) \, \frac{d\dot{M}}{d\Omega},
\end{equation}
where $g(p_{\mathrm{f}})\leq1$ is a function that 
accounts for the lack of spherical symmetry ($g<1$ for
$p_{\mathrm{f}}<R_{\mathrm{out}}$), and is estimated in
Appendix~\ref{appa}. For reference, if $f=1$,
$X_{\mathrm{OH}}^{-1} n_{\mathrm{OH}}=700$ cm$^{-3}$ at $r=70$ pc,
and $v=1000$ \kms, then
$d\dot{M}/d\Omega\approx90$ \Msun\ yr$^{-1}$ sr$^{-1}$. 
The momentum flux and the mechanical power, $\dot{M}v$ and $0.5\dot{M}v^2$,
are higher in this approach for the highest velocity gas.

The sizes we report below ($R_{\mathrm{int}}$, $R_{\mathrm{out}}$) should
be considered effective. Results identical to a given model are obtained by
scaling $R_{\mathrm{int}}$ and $R_{\mathrm{out}}$ to higher values as
$R/\sqrt{f}$ ($f<1$) while decreasing the densities as
  $n_{\mathrm{H}}\sqrt{f}$ (i.e. keeping the same column density) and
decreasing the continuum-subtracted spectra as $f\times F_{\nu}$. This would
{\em approximately} simulate partial covering by the outflow of the far-IR
source (covering factor $f$). Conversely, the emerging profiles can also
be interpreted as produced by an ensemble of $f$ clouds ($f>1$) each of
radius $R_{\mathrm{int}}/\sqrt{f}$. In both cases, the mass-outflow rate
scales as $\sqrt{f}$. A lower limit on $f$ is set by the condition that
  the modeled far-IR continuum cannot exceed the observed level.
We argue below (Sect.~\ref{sec:discussion} and \ref{sec:colhvc}) for nearly
complete covering ($f\sim0.4-1$) for both the high-excitation quiescent
component (QC) and the high-velocity component (HVC), and give below all
parameters ($R_{\mathrm{int}}$, $R_{\mathrm{out}}$, $n_{\mathrm{H}}$) for
$f=1$. For the low-excitation extended component (LEC) discussed below,
$f\approx0.2$ under the assumptions discussed in Sect.~\ref{sec:pcyg}.

   \begin{figure*}
   \centering
   \includegraphics[width=15.0cm]{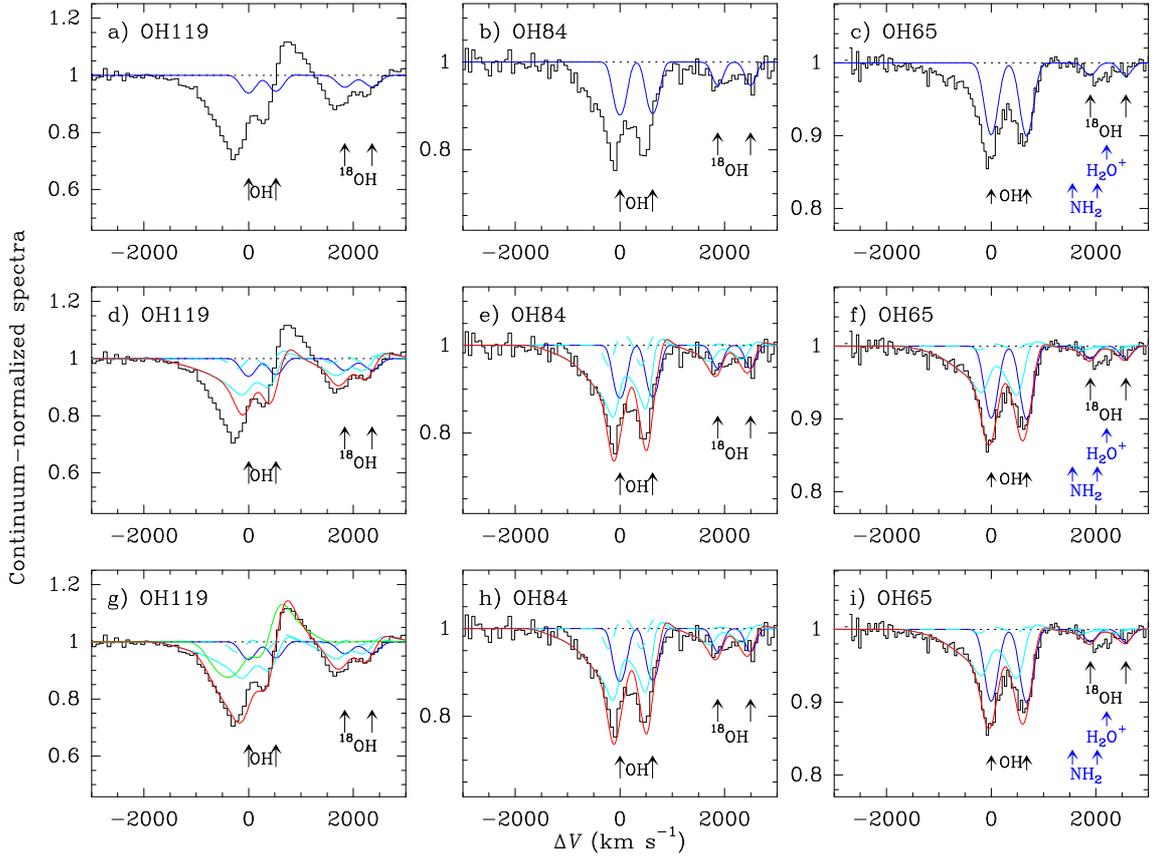}
   \caption{Illustration of the need for several OH components in Mrk~231, as
     inferred from the OH119, OH84, and OH65 doublets. In panels d-i, red
     curves indicate the absorption and emission by all considered
     components.  
     {\bf a-c)} An outflow-free, high-excitation component (QC) generates
     absorption in the high-lying OH lines (blue curves), but cannot account
     for the blue wings in the three doublets or the redshifted
     emission in the OH119 transition. {\bf e-f)} The HVC (light-blue curves)
     and LVC (dashed light-blue curves) reproduce the blue wings in the OH85
     and OH65 doublets, but fail to account for both the full
     blueshifted absorption and redshifted emission in the ground-state OH119
     doublet. {\bf g)} A low-excitation component (LEC, green curves) is 
     therefore 
     required to match the ground-state OH119 (and also OH79) blueshifted
     absorption and redshifted emission (panel g). The composite fit to all
     lines is shown in Fig.~\ref{fit}. 
   } 
    \label{components}
    \end{figure*}

We propose in this study a set of parameters for each component that provide
a reasonable match to the observed line profiles, though a more complete
study will be performed in combination with the other species detected within
the PACS domain. Generally speaking, the outflowing components that
are highly excited (as seen in OH) require compact sources (low
$R_{\mathrm{out}}$) and thus high mass-outflow rates.

   \begin{figure*}
   \centering
   \includegraphics[width=15.0cm]{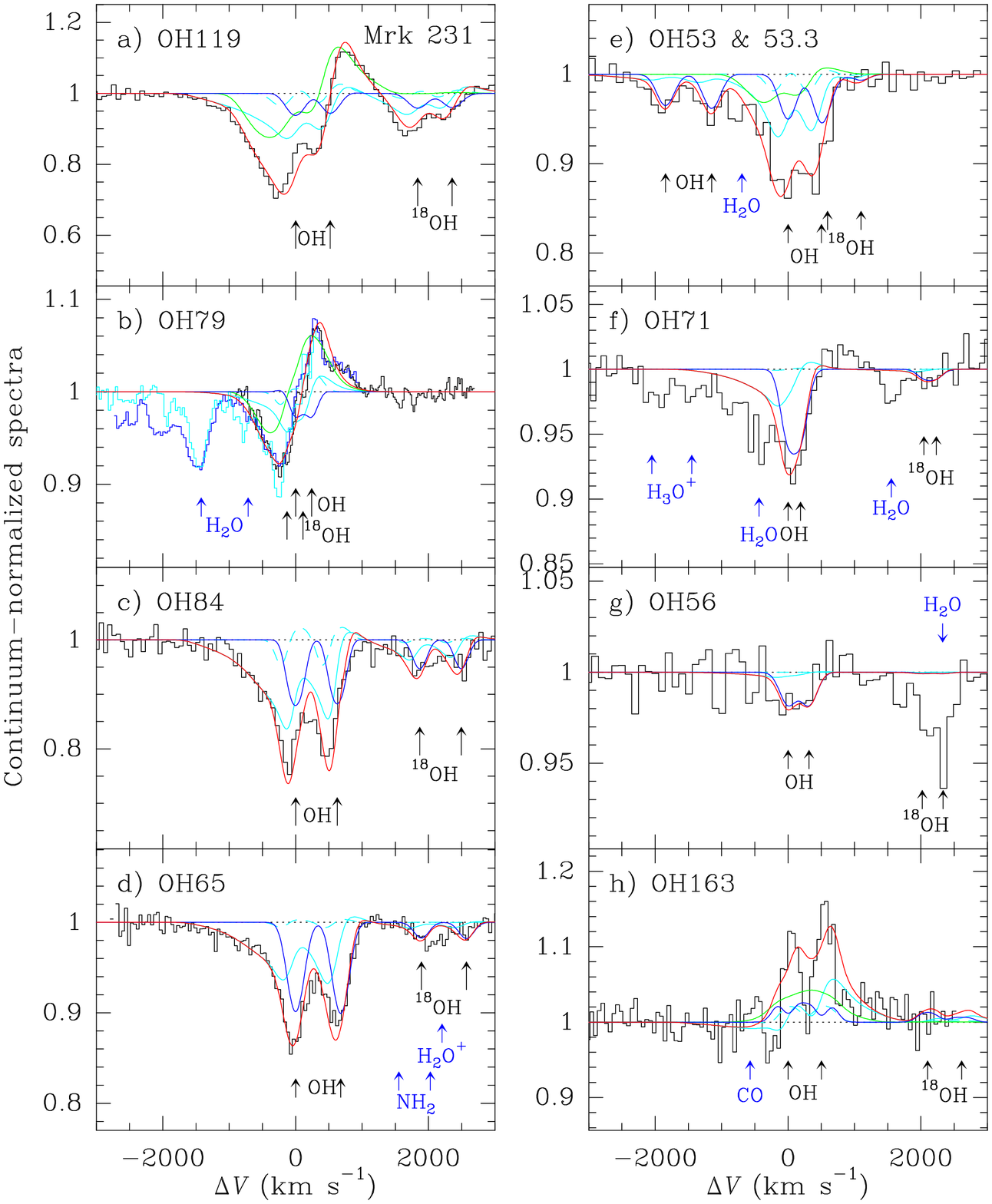}
   \caption{Model fit for the OH doublets in Mrk~231. The blue, light-blue,
     dashed light-blue, and green curves show the contributions by the
     quiescent component (QC), high-excitation outflow component (HVC),
     low-velocity component (LVC), and low-excitation component
     (LEC), respectively. Red is the total emission and absorption due to
       all components. In this specific model, the
       parameters for the HVC are 
     $T_{\mathrm{dust}}=105$ K, $R_{\mathrm{int}}=74$ pc, 
     $R_{\mathrm{out}}/R_{\mathrm{int}}=1.3$,
     $N_{\mathrm{OH}}=1.6\times10^{17}$ \cmd, 
     $v_{\mathrm{int}}=1700$ \kms, $v_{\mathrm{out}}=100$ \kms, and
     $p_{\mathrm{f}}=1.15\times R_{\mathrm{int}}$, and the mass-outflow rate
     per unit of solid angle is $d\dot{M}/d\Omega\approx100$ \Msun\ yr$^{-1}$
     sr$^{-1}$.
   } 
    \label{fit}
    \end{figure*}

\subsubsection{Gas components and the simplest model}
\label{sec:comp}

The need for several gas
  components is illustrated in Fig.~\ref{components} for the OH119, OH84, and
  OH65 doublets. Specifically, the OH84 and OH65 line
profiles (Fig.~\ref{mrk231}c-d) together with their ratio (Fig.~\ref{excit})
are primarily used to define the gas components where OH is excited, while an
additional low-excitation component is required to fully match the
ground-state OH119 and OH79 doublets. One of our best-fit composite
models is compared with all of the OH profiles in Fig.~\ref{fit}, where the
red curves indicate the total absorption and emission as generated from
all components.  

{\it -The quiescent component (QC)}: the spectra of the excited lines and
also the OH53.3 doublet indicate the presence of highly excited
gas with the lines peaking at central velocities. The model for the QC is shown
with blue curves in Figs.~\ref{components} and \ref{fit}. 

{\it -The high- and low-velocity components (HVC and LVC)}: in our
simplest model, most of the absorption in the blueshifted line wing of the
OH84 and OH65 excited doublets was simulated with a single outflowing
component with a negative velocity gradient, the HVC. This component is
indicated with light-blue curves in Figs.~\ref{components} and \ref{fit} and,
in addition to the wing in OH65 and OH84, it contributes significantly
to the absorption and emission in all other doublets, except for the
high-excitation OH53 and OH56. Details of our best-fit model for
the HVC are discussed and characterized in more detail in
Appendix~\ref{appa}.

The OH84 absorption at low velocities ($\sim200$ \kms) is not fully reproduced
with the HVC alone, and additional absorption at these velocities is proposed
to arise from another more extended low-velocity component (LVC). The LVC
contributes slightly to the reemission in OH119, OH79, OH84, and OH163.

{\it -The low-excitation component (LEC)}: the joint absorption and
emission from the above components fails to account for the 
total absorption and emission observed in the ground-state OH119 and OH79
doublets, therefore the additional LEC component (green curves in
Fig.~\ref{fit}) that produces significant absorption in these doublets but not
in excited lines is required. Because of the low excitation that this
component represents, it is also the most extended one, presumably tracing 
the outflow region probed at millimeter wavelengths by CO
and HCN. The parameters of this component are less constrained than
  those of the QC and HVC.

The inferred model parameters for the well-constrained components are
listed in Table~\ref{tab:parval}. We also list in Table~\ref{tab:ener} the
densities, hydrogen columns, and masses associated with the QC and the HVC,
as well as the energetics that characterize the HVC.

   \begin{table}
      \caption[]{Probable values of the parameters involved in the
        OH modeling$^{\mathrm{a}}$.} 
         \label{tab:parval}
          \begin{tabular}{lccc}   
            \hline
            \noalign{\smallskip}
Parameter  & QC & HVC & LVC$^{\mathrm{b}}$   \\  
            \noalign{\smallskip}
            \hline
            \noalign{\smallskip}
$R_{\mathrm{int}}$ (pc)$^{\mathrm{c}}$ & $55-73$ & $65-80$ & $65-80$   \\
$T_{\mathrm{dust}}$ (K) & $95-120$  & $90-105$ & $\sim90$  \\
$\tau_{100}$        & $1-3$ & $1.5-2.0$ & $\lesssim1$  \\
$R_{\mathrm{out}}/R_{\mathrm{int}}$ & $-$ & $\lesssim1.5$ & $\sim1.5-2$    \\
$v_{\mathrm{int}}$ (\kms) & $-$ & $1700$ & $\sim300$   \\
$v_{\mathrm{out}}$ (\kms) & $-$ & $100$ & $\sim200$   \\
$N_{\mathrm{OH}}$ ($10^{17}$ cm$^{-2}$) & $5-16$$^{\mathrm{d}}$ &
$1.5-3$ & $\sim0.3$  \\   
$p_{\mathrm{f}}/R_{\mathrm{out}}$$^{\mathrm{e}}$  & $1$  & $\sim0.8$ & $\sim1$  \\
            \noalign{\smallskip}
            \hline
         \end{tabular} 
\begin{list}{}{}
\item[$^{\mathrm{a}}$] Parameters for the LEC (low-excitation component)
are not well constrained (see Sect.~\ref{sec:lec}) and are omitted.
\item[$^{\mathrm{b}}$] Uncertain parameters from the present data. 
\item[$^{\mathrm{c}}$] For $f=1$.
\item[$^{\mathrm{d}}$] Column density per unit of $\tau_{50}$ (G-A12).
\item[$^{\mathrm{e}}$] $p_{\mathrm{f}}/R_{\mathrm{out}}=1$ is a fully
  spherical model, while $p_{\mathrm{f}}/R_{\mathrm{out}}<1$ simulates
  collimation in the direction of the observer.
\end{list}
   \end{table}

   \begin{table}
      \caption[]{Densities, column densities, masses, and energetics$^{\mathrm{a}}$} 
         \label{tab:ener}
          \begin{tabular}{lcc}   
            \hline
            \noalign{\smallskip}
Parameter  & QC & HVC    \\  
            \noalign{\smallskip}
            \hline
            \noalign{\smallskip}
$n_{\mathrm{H}}$ ($10^{4}$ \cmt)            & $1-2^{\mathrm{b}}$       & $0.04-0.3^{\mathrm{c}}$   \\
$N_{\mathrm{H}}$ ($10^{24}$ \cmd)           & $1.3-4$       & $0.06-0.12$  \\
$M_{\mathrm{gas}}$ ($10^{8}$ \Msun)          & $2.5-5.0$  & $0.2-0.4$   \\
$\dot{M}$ (\Msun\ yr$^{-1}$)              & $-$    & $500-1200$    \\
$\dot{P}$ (10$^{36}$ g cm s$^{-2}$) & $-$    & $\sim5-7^{\mathrm{d,e}}$    \\
$L_{\mathrm{mech}}$ (10$^{10}$ \Lsun)         & $-$    & $\sim6-10^{\mathrm{d,e}}$    \\
$T_{\mathrm{mech}}$ (10$^{56}$ erg)         & $-$    & $\sim2-4^{\mathrm{e}}$    \\
            \noalign{\smallskip}
            \hline
         \end{tabular} 
\begin{list}{}{}
\item[$^{\mathrm{a}}$] Assuming $X_{\mathrm{OH}}=2.5\times10^{-6}$ relative to
  H nuclei and $f=1$. Values scale inversely with the OH abundance
    relative to this assumed value. Only the best-constrained
    components, the QC and the HVC, are considered (see Sect.~\ref{sec:lec}
    for the LEC). 
\item[$^{\mathrm{b}}$] Average density (the medium is probably clumpy).
\item[$^{\mathrm{c}}$] The two values correspond to the highest and
  lowest outflowing velocities ($1700$ and $100$ \kms, respectively). 
\item[$^{\mathrm{d}}$] Varies with velocity; values are given for $v=1000$
  \kms.
\item[$^{\mathrm{e}}$] Values are given for $\dot{M}=850$
  \Msun\ yr$^{-1}$ and $R_{\mathrm{out}}/R_{\mathrm{int}}=1.3-1.5$.
\end{list}
\end{table}

\subsection{Quiescent Component (QC)}
\label{sec:qc}

The QC component is modeled as in G-A12; we adopt
a ``mixed'' approach (i.e. the OH molecules and the dust are
coexistent), and simulate the line broadening with a microturbulence
approach  ($v_{\mathrm{tur}}=90$ \kms). The line ratios depend on \tdust,
  $\tau_{100}$, and $N_{\mathrm{OH}}/\tau_{50}$ (G-A12), for which the
    explored ranges are $80-130$ K, $0.5-3$, and $(2-20)\times10^{17}$
    \cmd. In our best-fit model, the line ratios are reproduced with
  $T_{\mathrm{dust}}=110$ K, $\tau_{100}=1.5$, and an OH column of
  $8\times10^{17}$ \cmd\ per unit of $\tau_{50}$ (blue
curves in Fig.~\ref{fit}). For the above parameters, an effective ($f=1$)
radius of $R_{\mathrm{out}}\approx64$ pc is obtained. 
Similar model fits are also obtained by decreasing (increasing) \tdust,
and increasing (decreasing) both $N_{\mathrm{OH}}/\tau_{50}$ and
$R_{\mathrm{out}}$; the most plausible ranges are $T_{\mathrm{dust}}=95-125$ K,
$N_{\mathrm{OH}}/\tau_{50}=(16-5) \times10^{17}$ \cmd, and
$R_{\mathrm{out}}=73-55$ pc, respectively (Table~\ref{tab:parval}). For a
hydrogen column of $N_{\mathrm{H}}=4\times10^{23}$ \cmd\ per unit of
$\tau_{50}$ (G-A12), $N_{\mathrm{OH}}/\tau_{50}=8\times10^{17}$ \cmd\ gives
  $X_{\mathrm{OH}}=2\times10^{-6}$, in close agreement with our adopted value. 
Both $T_{\mathrm{dust}}$ and $N_{\mathrm{OH}}/\tau_{50}$ are similar to those
inferred in the nuclear region of Arp~220 (G-A12). The total OH
  column along a radial path (assuming uniform conditions) is
  $\sim4\times10^{18}$ \cmd. The continuum optical depth is similar to that of
  the HVC (Sect.~\ref{sec:hlvc}), $\tau_{100}=1.5$, and corresponds to 
  $N_{\mathrm{H}}\sim2\times10^{24}$ \cmd. Values of $\tau_{100}>3$
  would produce an absorption feature in the OH163 doublet at central
  velocities\footnote{The \hdo\ absorption lines, however, favor
    $\tau_{100}\sim3$, as will be reported in a future work.}, while
  $\tau_{100}<1$ would fail in reproducing the strong absorption in the OH71
  doublet at central velocities.

The QC is significantly smaller (for $f=1$) and most likely warmer than
the ``warm component'' needed to reproduce the 
H$_2$O lines detected in Mrk~231 with SPIRE ($95$ K and $120$ pc, 
  G-A10). 
Preliminary models for \hdo\ indicate, however, that the \hdo\ {\em
  absorption} lines are fitted with a model source size as small as that
of the QC. This behavior suggests that while the \hdo\ absorption is primarily
produced in front of the compact far-IR optically thick cores, efficient 
\hdo\ submillimeter line emission is generated in the surrounding more
extended regions with lower extinction, but with a far-IR
radiation density sufficiently high to produce the observed excitation. 
The relationship between the \hdo\ absorption and emission lines will be
explored in a future study.   

To avoid too strong absorption at central velocities in the
ground OH119 and OH79 doublets, significant collisional excitation with a
density of $2.5\times10^6$ cm$^{-3}$ (much higher than the average
  density of $(1-2)\times10^4$ \cmt, Table~\ref{tab:ener})
was used. As discussed in Sect.~\ref{sec:cenv}, however, significant
departures from spherical symmetry (i.e. reemission from a flattened and
possibly more extended structure viewed nearly face-on or with moderate
inclination) could also account for the weakness of these lines. Efficient
reemission at central velocities could take place in the surrounding region
primarily responsible for the \hdo\ submillimeter emission, or at even the
larger scales of the face-on disk \citep{dow98}. In any case, the above
density is effective because collisional excitation with electrons and
atomic hydrogen in an environment with a high-ionization fraction would
additionally relax the required value. Nevertheless, the density cannot be
lower than $\mathrm{several}\times10^4$ \cmt, indicating that the QC is
clumpy. In these effective models, collisional excitation 
primarily affects the ground-state OH119 and OH79 transitions.

\subsection{High- and low-velocity components (HVC and LVC)}
\label{sec:hlvc}

In our simplest approach, the observed absorption in the line wings of
the excited OH84 and OH65 doublets at $v<-400$ \kms\ are simulated with a
single outflow component, the HVC. It is characterized by a decelerating 
velocity field with $v_{\mathrm{int}}=1700$ \kms\ and $v_{\mathrm{out}}=100$ \kms,
with hydrogen column densities in velocity intervals of 100 \kms, as
shown in Fig.~\ref{cover}a.
We used two values for $T_{\mathrm{dust}}$; $105$ K, which is close to
the value used for the QC, and $90$ K, closer to the value used for the
  warm component in G-A10. A moderately high $\tau_{100}=1.5$ is favored,
motivated by the blueshifted absorption seen in the OH163 doublet.

\subsubsection{Column density and spatial scale of the HVC component}
\label{sec:colhvc}

   \begin{figure}
   \centering
   \includegraphics[width=8.0cm]{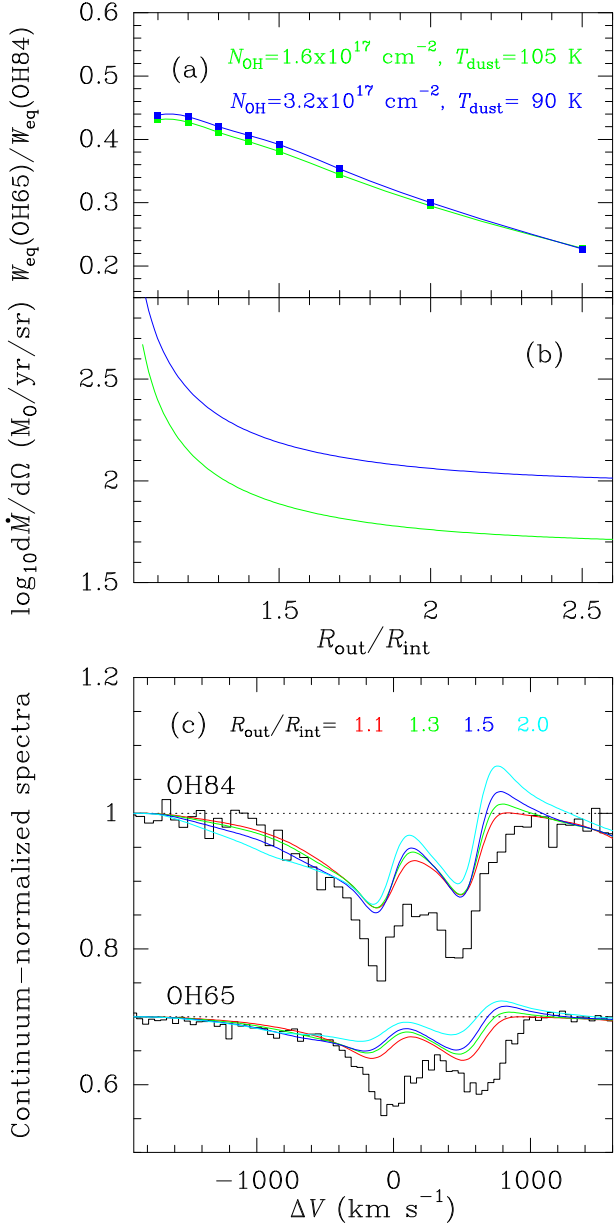}
   \caption{Model results for OH84 and OH65 in the HVC component. {\bf a)}
     The ratio of the OH65-to-OH84 equivalent widths in the blueshifted wing
     as a function of the thickness of the outflowing shell,
     $R_{\mathrm{out}}/R_{\mathrm{int}}$. $N_{\mathrm{OH}}$ and \tdust\ are
     given by $1.6\times10^{17}$ \cmd, $105$ K (green curve), and 
     $3.2\times10^{17}$ \cmd, $90$ K (blue curve), yielding very similar
     results. Other model parameters are $\tau_{100}= 1.6$,
     $v_{\mathrm{int}}=1700$ \kms, and $v_{\mathrm{out}}=100$ \kms. {\bf b)}
     Corresponding mass-outflow rates per unit solid angle for full
       coverage of the far-IR source ($f=1$, corresponding to
       $R_{\mathrm{int}}=75$ pc; see also eq.~\ref{eq:mdot2} and
       Table~\ref{tab:parval}). The assumed OH abundance relative to H
         nuclei is $X_{\mathrm{OH}}=2.5\times10^{-6}$. {\bf c)} 
     Observed (black histograms) and predicted OH84 and OH65 line profiles
     ($N_{\mathrm{OH}}=3.2\times10^{17}$ \cmd) for
     $R_{\mathrm{out}}/R_{\mathrm{int}}=1.1$ (red curves), $1.3$ 
     (green), $1.5$ (blue), and $2.0$ (light-blue).  
   } 
    \label{hvc}
    \end{figure}

How extended is the outflowing gas in the HVC (as seen in OH) as compared with
the source of far-IR emission that excites the OH? In Fig.~\ref{hvc}a, the
OH65-to-OH84 equivalent-width ratio (solid curves) is plotted
as a function of the adopted $R_{\mathrm{out}}/R_{\mathrm{int}}$ for two
combinations of (\tdust, $N_{\mathrm{OH}}$). The observed ratio
($\approx0.4-0.5$) can be reproduced either with ($T_{\mathrm{dust}}=105$ K,
$N_{\mathrm{OH}}=1.6\times10^{17}$ \cmd), or with ($T_{\mathrm{dust}}=90$ K,
$N_{\mathrm{OH}}=3.2\times10^{17}$ \cmd), as long as the thickness of the
outflowing shell is small in comparison to the radius of the far-IR
source (i.e., $R_{\mathrm{out}}/R_{\mathrm{int}}\lesssim1.5$). 

In a more extended outflow ($R_{\mathrm{out}}/R_{\mathrm{int}}>1.5$) the OH
becomes less excited, the predicted OH65/OH84 ratio drops, and higher columns
are then required. However, an extended HVC would have an observable effect on
the line shapes. In spherical symmetry, one would expect an emission feature at
redshifted velocities (see Fig.~\ref{hvc}c), arising from the limb of the
far-IR source where the continuum optical depth is relatively low (for
impact parameters $p$ higher than $R_{\mathrm{int}}$,
Fig.~\ref{outflow}). This modeled emission feature, especially prominent in the
OH84 doublet, is not seen in the spectra. Reemission in OH84 is not occurring
at {\em high} velocities, that is, not in the HVC (for the LVC, see
below), indicating that the projected surface where reemission by the excited
OH is generated is not significantly larger than the surface where the
absorption is produced. This suggests that either the HVC is compact around the
optically thick far-IR continuum source, or that the outflow is collimated
($p_{\mathrm{f}}\sim R_{\mathrm{int}}$, Fig.~\ref{outflow}), flowing just 
toward (and possibly in the opposite direction of) the observer.  
In an extended/collimated HVC, however, the line shapes would differ
significantly from the observations. The covering factor as a function
of the line-of-sight velocity would have little contrast between  
moderately low and high velocities, thus predicting relatively flat
blueshifted line wings that are hardly compatible with the observed steep OH84
blueshifted wing. The model fit for an extended outflow grossly overpredicts the
OH84 absorption at $\sim-1000$ \kms\ relative to OH65 (light-blue curve in
Fig.~\ref{hvc}b). While some degree of collimation is probably present 
(see Sect.~\ref{sec:pcyg} and below), the
observed line shapes and high OH excitation argue in favor of a component
where the high-velocity OH gas is piled up into a relatively narrow region,
tracing excited gas blowing out (along with the warm dust) from the warm
far-IR continuum source against which we see the OH absorption. We 
  therefore favor $R_{\mathrm{out}}/R_{\mathrm{int}}\lesssim1.5$. 

For the case of full coverage of the far-IR source ($f=1$), the size of
the far-IR source required to reproduce the absolute fluxes is 
$R_{\mathrm{int}}\sim65-80$ pc, and the outflow size (diameter) is
$\sim200$ pc. A lower limit, $f\gtrsim0.45$, is set by the constraint
  that the continuum flux density at $30-50$ $\mu$m, $7.5-12.5$ Jy,
  cannot exceed the observed continuum level, implying a physical size not
  higher than $1.5\times R_{\mathrm{int}}$ and an outflow size of up to
  $\sim300$ pc.  
The size of the far-IR continuum source behind the HVC is slightly larger 
 than but similar to that of the QC component, suggesting that the
outflow fully covers the QC.

The OH84 and OH65 blueshifted wings can be reproduced almost equally well with
($T_{\mathrm{dust}}=105$ K, $N_{\mathrm{OH}}=1.6\times10^{17}$ \cmd), and with
($T_{\mathrm{dust}}=90$ K, $N_{\mathrm{OH}}=3.2\times10^{17}$ \cmd), 
  illustrative of degeneracies in the models when constrained only by these
  two transitions. However, 
significant differences between the two models are seen especially in the
ground-state OH79 and OH53.3 doublets, which are overpredicted by the high
column-density solution. On the other hand, the high strength of the
$^{18}$OH85 doublet favors high OH columns (Sect.~\ref{sec:18oh}), so that the
column density is probably within the range
$N_{\mathrm{OH}}=(1.5-3.0)\times10^{17}$ \cmd. 

Even with a compact shell with $R_{\mathrm{out}}/R_{\mathrm{int}}=1.3$,
our spherical models overpredict the reemission at redshifted velocities in the
OH84 doublet profile, so that $p_{\mathrm{f}}/R_{\mathrm{out}}<1$ is favored
($\sim0.8$, Table~\ref{tab:parval}). As a
consequence of the compactness and collimation, the predicted redshifted
reemission in the HVC component of the OH119 and OH79 doublets is weak (see
also Fig.~\ref{components}), so the observed emission features remain as
residuals, and we attribute them to more spatially extended components,
that is, the LVC and mostly the LEC discussed below.

\subsubsection{LVC component}
\label{sec:lvc}

In most of the generated models for the HVC, the OH84 absorption at low
blueshifted velocities ($200-300$ \kms) is underpredicted. Broadening of the
absorption by the QC due to rotation of the circumnuclear structure 
(torus or thick disk)
could account for some of this missing absorption. However, the responsible
gas is less excited than in the HVC, as little additional absorption in
OH65 is required for a good fit of the profile. Therefore, we tentatively
associate this absorption with an increase of the covering factor of the
continuum by the OH at these velocities. Even if this additional
low-velocity 
absorption is probably produced by a spatial extension of the HVC in the plane
of sky, with velocities lower than predicted by the HVC, it is modeled in our
spherically symmetric models by means of a separate component, the LVC (dashed
light-blue lines in Fig.~\ref{fit}). The 
LVC is more extended than the HVC, generating some reemission in OH119,
OH79, OH84, and OH163. In general, the inferred parameters of the LVC are
rather uncertain because the associated absorption overlaps with that
  produced by the HVC; we modeled it
with $R_{\mathrm{out}}=150$ pc and $N_{\mathrm{OH}}=3\times10^{16}$ \cmd.
This is a minor component, contributing little to the observed
  spectra and only at low velocities.

\subsubsection{Velocity field?}
\label{sec:vfield}

The $\mathrm{OH65/OH84}$ ratio in the blueshifted line wing is relatively flat
for $v>-900$ \kms, and tends to increase (or at least remain similar) with 
higher velocity shifts. This dependence provides clues about the relative
location of the gas at different velocities with respect to the source of
excitation. If the OH excitation were independent of velocity, 
saturation of the OH84 doublet at low velocities (Fig.~\ref{cover}c) would
enhance the $\mathrm{OH65/OH84}$ ratio at these velocities. This is contrary
to the observed trend, suggesting that the OH gas with the highest velocity
shift is more excited than the low-velocity outflowing gas. The
increasing excitation with increasing velocity shift is, in our models,
generated by locating the higher velocity gas closer to the far-IR exciting
source than the lower velocity gas (Fig.~\ref{cover}b), thus suggesting an
overall decelerating velocity field. We also tried to model the HVC with
  accelerated velocity flows, but found that the modeled line shapes and line
  flux ratios were inconsistent with observations.
In our models, the LVC is more extended and 
less excited than the HVC, supporting the same decelerating
scenario. We note, however, that this solution relies on our simple spherical
geometry (where the successive shells are concentric) and may not be unique; 
for example, the high- and low-velocity gas may be flowing from
different regions of a circumnuclear torus or disk, characterized by
different \tdust\ and possibly with different projection effects
as the outflow widens. Nevertheless, some deceleration is 
most likely taking place because CO and HCN, which trace larger regions,
show wings up to a velocity of $\sim800$ \kms\ from the line center
\citep{fer10,cic12,aal12}, significantly lower than OH. \cite{spo09} also
inferred a decelerating velocity field from the ionized gas outflows traced by
the [NeII], [NeIII] and [NeV] lines in a sample of ULIRGs, though not in
Mrk~231; the fine-structure mid-IR lines trace an outflow
on a significantly smaller spatial scale, however.

The very strong velocity gradient used in our model fits, with the
gas velocities varying from $1700$ to $100$ \kms\ in a relatively short path
($\lesssim40$ pc), may be indicative of high clumpiness and
  turbulence within the flow, but also favors a nonconcentric origin of
  gas at different velocities. Nevertheless, strong shocks in swept-up
    gas of high density and column could in principle produce a 
strong deceleration of the previously accelerated gas. It is also worth
  noting that the LEC described below also indicates the presence of 
high-velocity gas (up to $\sim900$ \kms) detached from the nuclear
region, representing high-velocity gas that escapes from the nuclear
region along paths of least resistance.

\subsubsection{Energetics}
\label{sec:mdot}

For $N_{\mathrm{OH}}=1.5\times10^{17}$ \cmd, $R_{\mathrm{int}}=70$
pc, and $R_{\mathrm{out}}/R_{\mathrm{int}}\lesssim1.5$, the mass-outflow
rate per solid angle in the direction of the observer 
(eq.~\ref{eq:mdot2}) associated with the HVC is
$d\dot{M}/d\Omega\gtrsim 68 \,
\sqrt{f}\,(2.5\times10^{-6}/X_{\mathrm{OH}})$ \Msun\  yr$^{-1}$ sr$^{-1}$.
In spherical symmetry, this corresponds to
$\dot{M}\gtrsim850\,g(p_{\mathrm f})\sqrt{f}$ \Msun\ yr$^{-1}$,
but we favor a collimated outflow 
($p_{\mathrm f}\sim R_{\mathrm{int}}$) such that, for
$R_{\mathrm{out}}/R_{\mathrm{int}}=1.5$, the anisotropy function $g(p_{\mathrm
  f})$ may be as low as $\sim0.4$ (increasing steeply for smaller sizes). At
least $\mathrm{several}\times10^2$ \Msun\ yr$^{-1}$, and possibly $\sim10^3$
\Msun\ yr$^{-1}$, are inferred {\em locally} in the circumnuclear region of
Mrk~231 (Table~\ref{tab:ener}). However, it is just the compact nature of the
HVC gas that may suggest a non-steady flow, leaving open the possibility
of intermittency.

In our prescription, the momentum flux increases with gas velocity and is
given by $\dot{P}=5.3\times10^{36} (\dot{M}/850\,\mathrm{M_{\odot}\, yr^{-1}})
(v/10^3\,\mathrm{km\, s^{-1}})\,(2.5\times10^{-6}/X_{\mathrm{OH}})$
$\mathrm{g\,cm/s^2}$, or $\sim 15 \, L_{\mathrm{AGN}}/c$, 
adopting $L_{\mathrm{AGN}} = 2.8 \times 10^{12}$ \Lsun\
  \citep{vei09}. The corresponding mechanical luminosity is 
$L_{\mathrm{mech}}\sim6\times10^{10}$ \Lsun. The uncertainty in these
parameters ($\dot{M}$, $\dot{P}$, and $L_{\mathrm{mech}}$) is as
high as a factor $\sim3$ mainly because of geometry effects and
the uncertainty in the OH abundance. While our estimates for the rates 
($\dot{M}$, $\dot{P}$, and $L_{\mathrm{mech}}$) are roughly consistent with
those inferred from CO by \cite{fer10}, our integral values
($M_{\mathrm{gas}}$ and $T_{\mathrm{mech}}$) are much lower due to the
compactness of the HVC.

\subsection{Low-excitation component (LEC)}
\label{sec:lec}

While the joint emission/absorption from the above three (QC, HVC, and
LVC) components properly describes the observed absorption in the {\em
  excited} doublets, the ground-state OH119 and OH79 lines remain
underpredicted. An additional low-excitation component (LEC) that
accounts for the remaining 
OH119 and OH79 flux, but does not significantly contribute to the
excited OH doublets, was therefore included in the model. 
The LEC is expected to be more spatially extended than the source of
far-IR emission so that the OH molecules remain essentially in the
ground-state, and is also expected to be primarily responsible for the
emission features detected in OH119 and OH79 at redshifted velocities. 
Because this component is traced by the ground-state
doublets, no additional constraints on the spatial extent can be
inferred from the OH data. Nevertheless, it is reasonable to assume that
the LEC probes the relatively extended outflowing emission measured at
millimeter wavelengths \citep{fer10,cic12,aal12}. 

   \begin{figure}
   \centering
   \includegraphics[width=8.0cm]{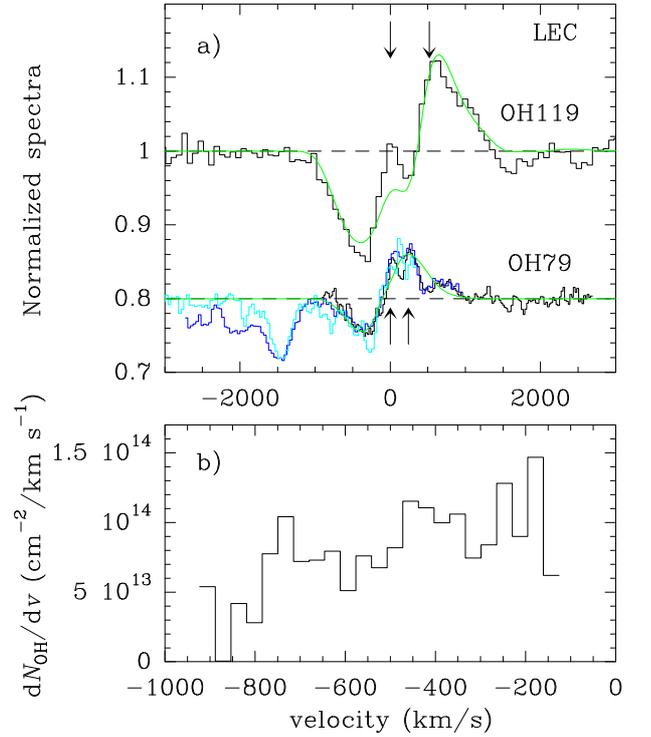}
   \caption{{\bf a)} OH119 and OH79 doublets after subtracting the model
     for the QC+HVC+LVC, thus tentatively isolating the contribution of the
     low-excitation component (LEC) to the absorption and emission. The
     green curves show our simple spherically symmetric model for the LEC
     (Sect.~\ref{sec:lec}). {\bf b)} 
     Inferred OH column density of the LEC per unit of line-of-sight velocity
     interval across the blue absorption wing after correcting for the
     covering factor at each velocity, but not corrected for the
    reemission in the lines (see text). The integral gives a total LEC column
     of $N_{\mathrm{OH}}\approx7\times10^{16}$ \cmd, in agreement with 
       the detailed models.  
   } 
    \label{lec}
    \end{figure}

Figure~\ref{lec}a shows the OH119 and OH79 profiles after subtracting the
  modeled emission of the QC+HVC+LVC (i.e. the modeled components for the
  excited OH). If our composite model for the excited OH is sufficiently
  accurate, the profiles in Fig.~\ref{lec}a thus isolate the contribution by
  the LEC to the observed profile. We note, however, that the
  emission at $v\sim0-200$ \kms\ may still have a substantial
  contribution from circumnuclear gas.
  In OH119, even collisional excitation in a warm, dense region may take place
  at  these moderately redshifted velocities where the CO 16-15 line appears to
  peak (Fig.~\ref{fit}h)\footnote{The ground-state lines of OH$^+$,
    CH$^+$, and HF are all detected in emission \citep{wer10}, indicating the
    importance of collisional excitation in these ground transitions; the
    observed emission in OH119 at central velocities may also have a substantial
    contribution from collisionally excited gas in the same warm/dense
    region.}.  
  The LEC contribution in Fig.~\ref{lec}a is thus tentative at low
    velocities. It is nevertheless 
  interesting that the OH119/LEC shows a nearly symmetric line shape with an
  emission feature only $\sim20\%$ weaker than the absorption feature, and
  with similar velocity extents on the blue and red sides. Within the model
  uncertainties and according to the discussion in Sect.~\ref{sec:pcyg}, this
  result is consistent with a roughly spherical distribution of the LEC with
  negligible extinction effects at 119 $\mu$m, indicating a wide opening
  angle of the flow at the corresponding spatial scales.   

Since detection of OH79 in the LEC ensures that the OH119 doublet is
  optically thick, the absorption of the LEC OH119 normalized spectrum 
  directly gives the covering factor at each line-of-sight blueshifted
  velocity ($f_v=1-F_v/F_c$, where $F_v/F_c$ is the continuum-normalized
  spectrum in Fig.~\ref{lec}a), uncorrected for the reemission in the
  line. The OH column per unit of velocity interval was estimated
  from the OH79 doublet (also uncorrected for the line reemission), and is
  shown in Fig.~\ref{lec}b. The integral of this spectrum gives
  $N_{\mathrm{OH}}\approx7\times10^{16}$ \cmd, in agreement with the model for the
  LEC discussed below that accurately takes into account the reemission in
  both doublets.

Models for the LEC have significant degeneracies because of
$(i)$ the uncertainty in the shape and strength of the far-IR continuum field
as seen by the absorbing and emitting OH, and $(ii)$ the lack of
constraints on the spatial scale. Our simple model for the LEC (green curves in
Fig.~\ref{lec}a and Fig.~\ref{fit}a,b,e, and h) assume the following:
$(i)$ the LEC surrounds the whole source of far-IR emission, which is
described by a spherical source with $R_{\mathrm{int}}=490$ pc,
$T_{\mathrm{dust}}=55$ K, and $\tau_{100}=0.5$ to nearly fit the observed
continuum between 50 and 130 $\mu$m; 
$(ii)$ we adopted an external radius of $R_{\mathrm{out}}=800$ pc
\citep[corresponding to the $\sim e^{-1}$ level of the $\mathrm{FWHM}=1.3$
kpc CO 1-0 line region,][]{cic12}\footnote{Note, however, that OH can
  potentially trace regions more extended than those traced by CO, because CO
  requires a minimum density to be collisionally excited while OH only needs the
  available far-IR radiation field.}. Finding a reasonable match
to the doublet shapes again requires a decelerating flow, with
$v_{\mathrm{int}}=900$ \kms\ and $v_{\mathrm{out}}=200$ \kms. The gas velocity
fields as derived from OH119 in other sources will be explored in a future
work.  

The model fit in Fig.~\ref{lec}a uses
  $N_{\mathrm{OH}}\approx6.3\times10^{16}$ \cmd\ (in close agreement with
  Fig.~\ref{lec}b), $p_{\mathrm{f}}=R_{\mathrm{out}}$
(i.e. strict spherical symmetry) and a covering factor of $f=0.20$ (as
discussed in Sect.~\ref{sec:pcyg}). The latter value is significantly lower
than the covering factor of the compact HVC ($f\gtrsim0.45$), possibly
indicating that the molecular outflow breaks into clumps as the gas expands
from the circumnuclear region. Indeed, the expected average density of
$n_{\mathrm{H}}\sim30$ \cmt\ is too low to excite the CO $1-0$
transition. The fit is reasonable except at central velocities in OH119,
suggesting further scattering in the doublet (see Sect.~\ref{sec:qc}). The LEC
carries most of the outflowing mass, 
$M_{\mathrm{gas}}\sim2\times10^8\times(2.5\times10^{-6}/X_{\mathrm{OH}})$
\Msun, and most of the mechanical energy,
$T_{\mathrm{mech}}\sim6\times10^{56}$ erg 
(compared with values of the HVC in Table~\ref{tab:ener}). Eq.~(\ref{eq:mdot2})
gives $\dot{M}\sim360\times(2.5\times10^{-6}/X_{\mathrm{OH}})$ \Msun\ 
yr$^{-1}$ for the above parameters. Within the uncertainties 
in the analysis of OH and CO \citep{fer10,cic12}, the energetics inferred from
both species appear to be consistent, especially if the OH abundance
drops below our adopted value at large distances from the circumnuclear
region.

\subsection{OH35 doublet}
\label{sec:oh35}

   \begin{figure}
   \centering
   \includegraphics[width=8.0cm]{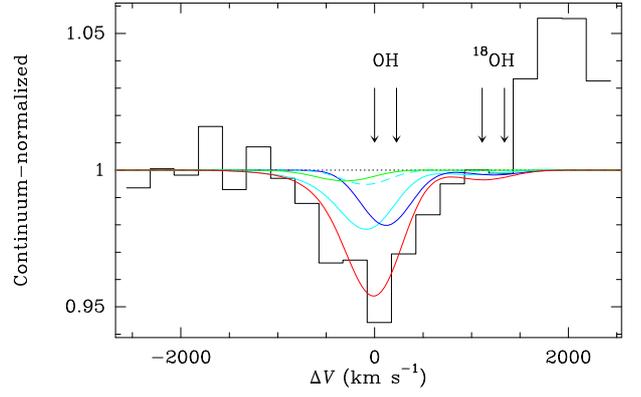}
   \caption{Model results for the OH35 doublet compared with the observed
     Spitzer IRS spectrum. The model and color code are the same as
     in Fig.~\ref{fit}.   
   } 
    \label{oh35fit}
    \end{figure}

In Fig.~\ref{oh35fit}, the same model used to fit the far-IR OH lines
observed with Herschel/PACS (Fig.~\ref{fit}) is compared with the Spitzer
IRS OH35 spectrum. While the absorption at central velocities is reproduced,
the model appears to underpredict the absorption at $\sim-500$ \kms\ as well
as at redshifted velocities. Since the ground-state OH53.3 is reproduced, these
discrepancies are most likely consequences of the uncertainties in the modeled
continuum-flux density at 35 $\mu$m relative to the flux density at longer
wavelengths, which is determined by the solid angle and \tdust\ of the
underlying continuum source. Specifically, we may expect a range of \tdust\
behind the observed absorption, with the warmest and most compact 
components contributing significantly to the mid-IR continuum emission.

\subsection{$^{18}$OH}
\label{sec:18oh}

One intringuing finding in the OH spectra of Mrk~231 is the
relatively strong absorption by $^{18}$OH seen at 120, 85, and 66
$\mu$m. While the $^{18}$OH120 doublet may be contaminated by CH$^+$ in its
blue component and the $^{18}$OH66 feature has a probable contribution by
NH$_2$ and H$_2$O$^+$, the prominent $^{18}$OH85 is expected to be free
from contamination and shows evidence for absorption by outflowing gas
as well as by the QC component. 

In our model for the HVC with $T_{\mathrm{dust}}=105$ K, we required
$N_{\mathrm{^{18}OH}}\sim1\times10^{16}$ \cmd\ to generate the modeled
blueshifted absorption in the $^{18}$OH85 doublet (Fig.~\ref{fit}c).
Likewise, $N_{\mathrm{^{18}OH}}\sim1.5\times10^{16}$ \cmd\ was obtained for
$T_{\mathrm{dust}}=90$ K, corresponding to
$\mathrm{OH/^{18}OH}\sim20$. Similarly, we required for the QC 
$N_{\mathrm{^{18}OH}}\approx3\times10^{16}$ \cmd\ per unit of $\tau_{50}$, 
that is, $\mathrm{OH/^{18}OH}\sim30$. The overabundance estimated for
$^{18}$OH is then even more extreme than we previously reported (F10). 
Since we cannot exclude higher OH columns at moderate velocities in the
outflow (because of saturation in OH84, Fig.~\ref{cover}c), we favor
$\mathrm{OH/^{18}OH}\sim20-30$ in both components, with some 
indications that the ratio decreases in the HVC. Models for the
  undetected $^{17}$OH were also performed, from which we estimate
  $\mathrm{^{18}OH/^{17}OH}\gtrsim5$.

\subsection{More details of the model fit and discrepancies}

There are several spectral features that our modeling does not account
for. The high-velocity redshifted emission wing in OH79 is poorly
reproduced, and the redshifted emission feature in OH71 is ignored. 
The latter may be associated with outflowing gas more excited than modeled for
the HVC.

The OH163 is one of the most puzzling of the line shapes
(Fig.~\ref{fit}h). The strength of the emission and absorption in this
doublet is very sensitive to the continuum opacity. The absorption at
blueshifted velocities and the asymmetry between the two lambda-doubling
components indicate high continuum opacity and thus suggest a significant
contribution by the HVC. However, the narrow linewidths of
the emission features would suggest an origin in low-velocity gas, but
both the QC and the LEC predict line shapes broader than observed.
The dip of emission in between the two lambda-doubling components cannot be
reproduced. Since the OH163 doublet is pumped through absorption
  of far-IR photons, part of the emission is most likely arising from
  the same region that generates the submillimeter \hdo\ emission (G-A10),
  which is expected to surround the QC (see Sect.~\ref{sec:qc}).

\section{Discussion and conclusions}
\label{sec:discussion}

\subsection{Summary}

The picture that emerges from the OH observations and models can be summarized
as follows: a highly excited component where OH peaks at central
velocities, the QC, represents an outflow-free circumnuclear
component with $T_{\mathrm{dust}}\sim110$ K, an effective radius $R\sim65$
pc, and a column of $N_{\mathrm{H}}\sim2\times10^{24}$ \cmd. The observed
high-velocity absorption by excited OH arises from a somewhat larger
($R_{\mathrm{int}}\sim75$ pc, $R_{\mathrm{out}}\sim100$ pc,
both effective radii) radiatively excited and apparently
  collimated component (the HVC).  
This component is also associated with high far-IR radiation density
($T_{\mathrm{dust}}\sim100$ K) and, given its somewhat larger size,
most likely surrounds the QC. This scenario suggests that the QC is
feeding the outflow, in the sense that the outflowing gas emanates from the
same circumnuclear structure that is responsible for the central-velocity
absorption. The OH column density in the 
HVC is $N_{\mathrm{OH}}\approx(1.5-3)\times10^{17}$ \cmd, suggesting $A_v\sim 30$
magnitudes of outflowing circumnuclear gas. We estimate a mass-outflow 
rate per unit of solid angle in the direction of the observer of at least
$\sim70$ and possibly $\sim100$ \Msun\ $\mathrm{yr^{-1}\,sr^{-1}}$ for
$X_{\mathrm{OH}}=2.5\times10^{-6}$. In spherical symmetry, 
this would correspond to $\sim1000$ \Msun\ $\mathrm{yr^{-1}}$, though
significant departures from a fully spherical model probably reduce the
above value by a factor $\sim2$. The 
momentum flux attains $\sim15\,L_{\mathrm{AGN}}/c$. In our models, 
  consisting of concentric shells of gas and dust with 
well-ordered radial motions, the high excitation found for the highest
velocity gas was reproduced 
with a decelerating flow (see discussion in Sect.~\ref{sec:vfield}). 
An extraordinary enhancement of $^{18}$OH was found
($\mathrm{OH/^{18}OH}\lesssim30$) in both the QC and the HVC. 

Our model for the excited OH leaves residuals in the ground-state OH119
  and OH79 doublets, indicating the presence of a low-excitation
  component of 
  the outflow (the LEC), with a column of $N_{\mathrm{OH}}\approx7\times10^{16}$
  \cmd. The LEC contribution to the profiles (Fig.~\ref{lec}a) is only
  tentative at low redshifted velocities, but appears to show similar
  strengths and velocity extents for the absorption and emission
  features in OH119. This suggests that the LEC is roughly spherical and
  spatially extended, in contrast with the HVC. If the LEC is extended
  and surrounds the whole source of 119 $\mu$m continuum 
  emission, its covering factor is $f\sim20$\%, significantly lower than that
  of the compact HVC ($f\gtrsim45$\%), possibly indicating that the
  molecular outflowing gas breaks into clumps as it moves away from the
  circumnuclear region. If some of the OH119 emission is
  circumnuclear and/or collisionally excited, the covering factor of the
  extended component must be even lower than $\sim20$\%.

\subsection{Torus or thick disk}
\label{sec:torus}

The QC has a modeled size (diameter of $\sim130$ pc) remarkably
similar to that of the circumnuclear rotating structure 
(torus or thick disk) observed with the EVN in OH megamaser emission by
\cite{klo03}, which delineates the central region of the OH megamaser
complex \citep{ric05} and traces the inner region of the radio/H I disk
\citep{car98} and of the star-forming region observed in the near-IR
\citep{dav04,dav07}. From the (roughly) estimated continuum optical depth
($\tau_{50}\sim5.5$) and size, the gas mass of the QC is
$\sim3\times10^8\times(0.01/X_{\mathrm{dust}})$ \Msun\  
($X_{\mathrm{dust}}$ is the dust-to-gas mass ratio)\footnote{This gives a
    gas mass surface density of $2\times10^4$ \Msun\ pc$^{-2}$, which
    is a lower limit to the total value including stars
    \citep{dav04,dav07}.}, 
significantly higher than the previously estimated virial mass
\citep{klo03}, but still roughly consistent within the uncertainties of
both estimates. Furthermore, our inferred 
$T_{\mathrm{dust}}\sim110$ K is not far from the value calculated via  
$(L_{\mathrm{AGN}}/4\pi R^2 \sigma)^{1/4}\sim130$ K, where
$L_{\mathrm{AGN}}\sim2\times10^{12}$ \Lsun\ (the actual \tdust\ will be
lower due to opacity effects). We thus tentatively identify the QC component
with the circumnuclear OH-megamaser torus (thick disk or oblate
spheroid geometries are equally favorable). The match in sizes also
suggests that $f$, the covering factor (Table~\ref{tab:par}), is of
order unity for the QC, though more likely $f\sim 0.4 - 1.0$ for both the
QC and HVC because the corresponding absorptions are surely produced
in different areas of the warm far-IR surface.   

The lack of OH119 absorption at central velocities may be indicative of
high densities in the quiescent component, but could also reflect
scattering (i.e. reemission in the line) taking place in a flattened structure
seen nearly face-on or with low inclination (Sect.~\ref{sec:cenv}). The
geometric problem is probably complex, because high resolution
observations indicate a tilt of the torus \citep{klo03,dav04,ric05}
relative to the outer nearly face-on disk \citep{dow98}; nevertheless, the
scattering process may be operating in the region responsible for the
  \hdo\ submillimeter emission (which is more extended than the QC, G-A10) or
  even on relatively large ($\sim1$ kpc) scales. Radiative-transfer
  models in 2D (axial symmetry) are required to distinguish between these
  scenarios.

\subsection{Radiatively excited molecular outflow}

The HVC is likely to be emanating from, or is at least associated
with, this torus, because the highly excited outflowing {\em absorbing}  
OH is seen in front of, and is excited by a strong far-IR radiation field
most likely generated in and around that circumnuclear component. 
The QC could also provide a reservoir of gas rich in OH that feeds the
  outflow,  but if so, then either a relatively smooth acceleration process
(e.g. successive low-velocity, non-dissociative C-shocks) allows the OH
to survive, or if the OH is destroyed in fast (J-) shocks, it must reform
in the post-shock gas.\footnote{The transition from a C- to a J-type
  shock occurs at a critical velocity of $2.76\,B/\sqrt{4\pi\rho_n}$
\citep[e.g.][]{cio04}, yielding $<100$ \kms\ for $B\sim300$ $\mu$G
\citep{car98} and a density of $10^3$ \cmt, much lower than the OH
velocities.} There is evidence for interaction between the radio jet and the 
surrounding gas \citep{ulv99,klo03,rup11}, as well as an overall (moderate)
velocity blueshift of the torus or thick inner disk relative to the
surrounding gas at larger spatial scales \citep{klo03} that could
  indicate a slow expansion of the torus. According to 
outflow models driven by radiation pressure \citep{rot12}, the low outward
velocity of the gas in the torus can be a result of high inertia, the drop of
the radiation pressure with decreasing \tdust, and gravitation.

The geometry of the inner outflowing gas (HVC) relative to the
torus may be more complex than simulated in our schematic spherically
symmetric models.  The possibility that the molecular gas is primarily flowing
along the polar regions of the torus has two drawbacks; first, the tilt
of the torus implies that its axis deviates from the direction of the
observer, with the consequent projection effects on the line-of-sight velocity
of the polar gas. Second, the gas column along the polar direction is expected
to be relatively low, while our inferred high mass-outflow rate and the
requirement of absorption of and excitation by optically thick 84/65 $\mu$m
continuum indicate large gas reservoirs behind (and associated with) the
outflowing gas. The 3D radiation pressure models by \cite{rot12}
predict the highest differential mass-outflow rates for polar angles
$>45^{\circ}$ (their Figs.~12-14), that is, not far from the equatorial
plane, and it is just the tilt of the torus that in this context would
  provide a geometry favorable for detecting high differential
  mass-outflow rates in the 
direction of the observer. Conceivably, the observed OH outflow could probe an
interclump medium of the torus itself that is flowing past the dense clumps
(possibly probed by the QC), permeating the whole structure. Interaction with
the high-density clumps and shadowing effects \citep{rot12} would
decelerate the outflowing gas with increasing radial distance. The 
highest-velocity gas could also be tracing a conical transition region
between the torus and the polar directions. In our model for the HVC, the
densities for velocities of $500-1500$ \kms\ are $n_{\mathrm{H}}\sim1000-500$
\cmt, respectively, also in rough agreement with the wind-driven outflow
models by \cite{fau12}.

\subsection{Role of the AGN}

The high mass-outflow rate and outflow velocities derived from
the far-IR observations of OH strongly point toward a key role of the
central AGN, as previously argued (S11). The momentum flux of $\dot{P} \sim 15
\, L_{\mathrm{AGN}}/c$ is roughly consistent with that required to regulate the
growth of the black hole and set the $M_{\mathrm{BH}}-\sigma$ relation
\citep{deb12}. In the framework of radiation pressure, 3D models indicate that
in a clumpy disk with a wide opening angle, the radiation tends to escape
along the poles and radiation pressure becomes less efficient, generally
accounting for a momentum deposition rate of $(1-5) \, L_{\mathrm{AGN}}/c$
\citep{rot12}. Still, these models predict high differential 
  mass-outflow rates 
($d\dot{M}/d\Omega>30$ \Msun\ $\mathrm{yr^{-1}\,sr^{-1}}$) for sufficiently
high columns and in directions close to the equatorial
plane; a high scale-height of the torus/disk, or a relatively high 
mass-concentration in the polar region, could additionally increase 
the mass-outflow rate. In addition, fast energy-conserving AGN winds can
do work on the swept-up (molecular) gas and then strongly boost the momentum
flux \citep{fau12}. It is 
possible that while radiation pressure affects the whole circumnuclear
structure, winds are responsible for the highest velocity wings seen in OH.
On the other hand, the high rate of mass loss derived here, together with the
narrow-shell configuration favored for the HVC component, may suggest an
intermittent (explosive) instead of a steady flow, consistent with the
multiple, expanding, concentric supershells seen in the optical/UV at larger
scales \citep{lip05,lip09}.

\subsection{$^{18}OH$ and the circumnuclear star formation}

An intriguing implication of the present observations is the strong
enhancement of $^{18}$OH in both the QC and the HVC. Since 
fractionation effects do not chemically enhance 
$^{18}$OH \citep{lan84}, the $^{16}$OH/$^{18}$OH ratio is expected to
be the same as the $^{16}$O/$^{18}$O ratio. Our results indicate that
$^{18}$O is enhanced by about one order of magnitude relative to the
Galactic Sgr B2 \citep{pol05}, and even more relative to the solar value. 
This is of interest in the context of the high metallicities 
that are observed in quasar environments, whose enrichment is thought to be
due to star formation with an IMF weighted toward massive stars
\citep[see review by][]{ham07}. Similarly, $^{18}$O is thought to be enriched
in the ISM by partial He burning in massive stars
\citep[e.g.][]{wil92,hen93,pra96,wou08,kob11}, 
and the stars in the inner disk of Mrk~231 have been formed in situ
\citep{dav04}. This is consistent with the lack of detection of $^{17}$OH 
if $^{17}$O is primarily produced in low- and intermediate-mass stars
  \citep{sag91,wil92} that, regardless of the IMF, are not expected to have a
significant chemical effect on the ISM of Mrk~231 \citep{mul06} given
the youth of the circumnuclear starburst, $\lesssim0.25$
Gyr \citep{dav07}. \cite{art93} and \cite{col99} have 
proposed and explored a scenario in which unstable fragments trapped in the
accretion disk of massive black holes, grow by accretion to $\sim50-100$
\Msun\ stars and ultimately explode as supernovae (SNe), to explain 
the metal enrichment of QSO ejecta, though it is unclear whether the
  $^{18}$O would be destroyed in these conditions by He burning, yielding
  $^{22}$Ne \citep[e.g.][]{pra96}. Interestingly, 
a very low (but not so extreme) $\mathrm{^{16}O/^{18}O}\sim50$ ratio,
together with a high $\mathrm{^{18}O/^{17}O}\sim12$ ratio, are also 
inferred in the arm of a spiral galaxy at $z=0.89$ \citep{mul06,mul11}.
In Mrk~231, $^{18}$OH is detected up to a velocity shift
of $\sim-600$ \kms, although $^{18}$OH enhancement at higher velocities is not
ruled out. While the high column density of $^{18}$OH in
the QC indicates previous $^{18}$O enrichment of the swept-up
  gas\footnote{Scaling the results by \cite{dav07} for
  an estimated SN rate of $\sim4$ yr$^{-1}$ within the inner $\sim500$ pc 
  \citep{dav04}, the cumulative ejected mass is $\sim2\times10^9$ \Msun\
  \citep[including OB winds and AGB stars,][]{dav07}, which is similar to
  the current stellar mass \citep{dav04} and also similar to the
  total gas mass \citep[$\sim1.8\times10^9$ \Msun,][]{dow98}. If a significant
  fraction of these ejecta still remains bound, the circumnuclear ISM is
  expected from these grounds to be deeply recycled by the ejecta of 
  high-mass stars.},    
the possible relative enhancement of $^{18}$OH that we inferred in the
line wing could suggest the contribution of SNe or massive stellar winds to
the outflow. More studies of OH/$^{18}$OH in galaxies and SNe
are required to fully understand the evolutionary implications of these
enhancements.

\begin{acknowledgements}
PACS has been developed by a consortium of institutes
led by MPE (Germany) and including UVIE (Austria); KU Leuven, CSL, IMEC
(Belgium); CEA, LAM (France); MPIA (Germany); 
INAFIFSI/OAA/OAP/OAT, LENS, SISSA (Italy); IAC (Spain). This development
has been supported by the funding agencies BMVIT (Austria), ESA-PRODEX
(Belgium), CEA/CNES (France), DLR (Germany), ASI/INAF (Italy), and
CICYT/MCYT (Spain). E.G-A is a Research Associate at the Harvard-Smithsonian
Center for Astrophysics, and thanks the Spanish 
Ministerio de Econom\'{\i}a y Competitividad for support under projects
  AYA2010-21697-C05-0 and FIS2012-39162-C06-01. 
Basic research in IR astronomy at NRL is funded by 
the US ONR; J.F. and H.W.W.S. also acknowledge support from the NHSC. 
S.V. thanks NASA for partial support of this research via Research Support
Agreement RSA 1427277, support from a Senior NPP Award from NASA, and
his host institution, the Goddard Space Flight Center, and
acknowledges support from the Alexander von Humboldt Foundation
for a renewed visit to Germany following the original 2009 award.
This research has made use of NASA's Astrophysics Data System (ADS)
and of GILDAS software (http://www.iram.fr/IRAMFR/GILDAS).
\end{acknowledgements}

\begin{appendix}
\section{Some properties of the OH outflow models}
\label{appa}

In our models, which assume a constant mass-outflow rate and velocity
gradient within $R_{\mathrm{int}}$ and $R_{\mathrm{out}}$, the relationship between
the mass-outflow rate per unit of solid angle ($d\dot{M}/d\Omega$) and
$N_{\mathrm{OH}}$ is given by (from eqs.~\ref{eq:col} and \ref{eq:mdot})
\begin{eqnarray}
d\dot{M}/d\Omega & = &  f \, m_{\mathrm{H}} \, 
X_{\mathrm{OH}}^{-1} \, v_{\mathrm{int}} R_{\mathrm{int}} N_{\mathrm{OH}} 
\times \left( \frac{x- v_{\mathrm{out}}/v_{\mathrm{int}}}{x-1} \right)
\nonumber \\ 
& \times &
\left[ 1 - \frac{1}{x} + 
\frac{v_{\mathrm{int}}/v_{\mathrm{out}}-1}{v_{\mathrm{int}}x/v_{\mathrm{out}}-1}
\ln \left( \frac{v_{\mathrm{int}}x}{v_{\mathrm{out}}} \right) \right]^{-1},
\label{eq:mdot2}
\end{eqnarray}
where $x\equiv R_{\mathrm{out}}/R_{\mathrm{int}}$. For given
$N_{\mathrm{OH}}$, $R_{\mathrm{int}}$, $v_{\mathrm{int}}$, and
$v_{\mathrm{out}}$, both $\dot{M}$ and the OH excitation increase with
decreasing $x$ (Fig.~\ref{hvc}a), so that higher excitation (e.g. higher
OH65/OH84 ratio) implies a more compact outflow and an increasing
$\dot{M}$ in our models.

   \begin{figure}
   \centering
   \includegraphics[width=8.0cm]{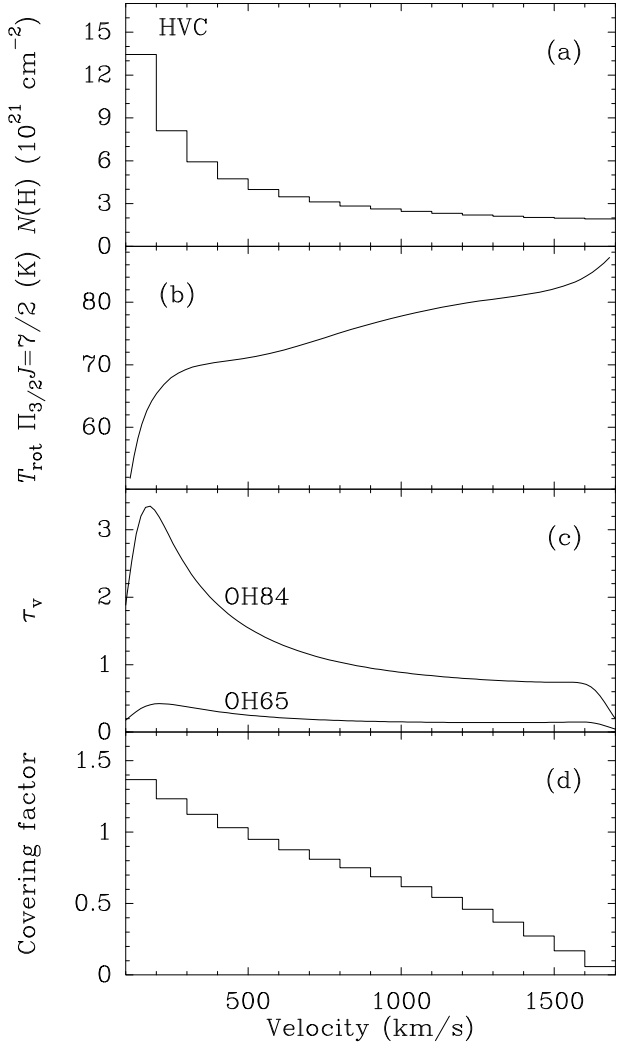}
   \caption{Details of the model for the HVC shown in Fig.~\ref{fit} with
     light-blue curves; note that higher velocities correspond to lower
     distances to the far-IR exciting source in a decelerating field. 
     {\bf a)} Column density of H nuclei (assuming  
     $X_{\mathrm{OH}}=2.5\times10^{-6}$) in intervals of 100 \kms\ as a
     function of the {\em radial} velocity for
     $R_{\mathrm{out}}/R_{\mathrm{int}}=1.3$, 
     $N_{\mathrm{OH}}=1.6\times10^{17}$ \cmd, $v_{\mathrm{int}}=1700$ \kms,
     and $v_{\mathrm{out}}=100$ \kms. {\bf b)} The rotational
     temperature of the $\Pi_{3/2}\,J=7/2$ level (the lower level of the OH65
     transition) relative to the ground state, as a function of the {\em
       radial} velocity. {\bf c)} The OH84 and OH65 optical depths along a ray
     passing through the center, and {\bf d)} the covering factor of
     the continuum, both as a function of the {\em line-of-sight}
     velocity. At low projected velocities ($<400$ \kms), the covering factor
     exceeds unity, which generates reemission from the limb of the outflow.
   } 
    \label{cover}
    \end{figure}

The OH column density per unit of velocity interval is
\begin{eqnarray}
\frac{dN_{\mathrm{OH}}}{dv} = 
\frac{\dot{M}\,
  X_{\mathrm{OH}}\,R_{\mathrm{int}}}{4\pi\,g(p_{\mathrm{f}})\,f\,
  m_{\mathrm{H}}} 
 \times \frac{x-1}{|v_{\mathrm{out}}-v_{\mathrm{int}}|} 
 \times \frac{1}{r^2 v(r)}.
\label{eq:dndv}
\end{eqnarray}
The corresponding $N_{\mathrm{H}}$ spectrum, calculated in velocity
intervals of 100 \kms, is shown in Fig.~\ref{cover}a for the model of the HVC
displayed in Fig.~\ref{fit}a. In a compact decelerating outflow, the
column density remains nearly constant for high velocities. The visual
extinction at high velocities in these 100 \kms\ intervals is expected to be
$A_V\sim1$ mag. 

The modeled line shapes depend on the velocity distribution of
$N_{\mathrm{OH}}$ and of the OH excitation, and on the covering factor as a
function of the line-of-sight velocity. The increasing excitation with
increasing velocity shifts is obtained in our models with a decelerating field
(Fig.~\ref{cover}b). The calculated OH84 and OH65 optical depths along a
radial path are shown in Fig.~\ref{cover}c, indicating saturation effects in
the OH84 doublet mostly at moderate velocities, but optically thin absorption
in OH65. On the other hand, the steep decrease of the OH84
absorption with increasing velocity shift (Fig.~\ref{fit}) is indicative of a
decreasing covering factor with increasing 
projected velocity, as shown in Fig.~\ref{cover}d. At low
projected velocities ($<400$ \kms), the covering factor exceeds unity,
which generates reemission from the limb of the outflow at significantly
redshifted velocities. Since this reemission is not observed in OH84, a
collimated ($p_{\mathrm{f}}\sim R_{\mathrm{int}}$) outflow is favored.

The total mass-outflow rate is given by
\begin{eqnarray}
\dot{M} = 4 \pi \, g(p_{\mathrm{f}}) \, d\dot{M}/d\Omega,
\label{eq:mdot3}
\end{eqnarray}
where $g=1$ for $p_{\mathrm{f}}=R_{\mathrm{out}}$. For 
$R_{\mathrm{int}}\leq p_{\mathrm{f}} < R_{\mathrm{out}}$, we roughly
approximate the geometry depicted in Fig.~\ref{outflow} as two cones, each
one with half opening angle
$\sin\theta_{1/2}=p_{\mathrm{f}}/R_{\mathrm{out}}$,
and thus 
\begin{eqnarray}
g(p_{\mathrm{f}})=1-\sqrt{1-(p_{\mathrm{f}}/R_{\mathrm{out}})^2}.
\end{eqnarray}
This approximation underestimates $\dot{M}$ because the model still includes
the contribution by gas outflowing along the plane of sky
(Fig.~\ref{outflow}).

\end{appendix}

\end{document}